\documentclass[10pt, conference]{IEEEtran}
\pdfoutput=1

\usepackage[activate={true,nocompatibility},final,tracking=true,kerning=true,spacing=true]{microtype}
\usepackage{siunitx}
\usepackage{gensymb}
\usepackage{openbib} 
\usepackage{tabularx,tabulary}
\usepackage{booktabs}
\usepackage{multirow}
\usepackage{amsmath,amssymb,amsfonts}
\usepackage{algorithmic}
\usepackage{graphicx}
\usepackage{textcomp}
\usepackage{enumitem}
\usepackage{xcolor}
\usepackage[hyphens]{url}
\usepackage[normalem]{ulem} 
\usepackage{wasysym}
\usepackage{xspace}
\usepackage{comment}
\usepackage{bibentry}
\usepackage{hyperref}
   \hypersetup{%
    unicode=true,%
    pdfstartview={FitH},%
    colorlinks=true,%
    citecolor=red,%
    linkcolor=red,%
    urlcolor=red,
    nolinks=false
}
\usepackage{comment}
\usepackage{graphicx}

\usepackage[numbers,sort&compress,square]{natbib}

\usepackage{subfig}
\usepackage{color}
\usepackage{icomma}

\let\oldcircled\textcircled
\renewcommand{\textcircled}[1]{\raisebox{.6pt}{\oldcircled{\raisebox{-.6pt}{#1}}}}

\newcommand{\comments}[1]{}

\newcommand{\fft}{FFT\xspace}
\newcommand{\xcor}{XCOR\xspace}
\newcommand{\bbf}{BBF\xspace}
\newcommand{\svm}{SVM\xspace}
\newcommand{\svms}{SVMs\xspace}
\newcommand{\thr}{THR\xspace}
\newcommand{\hconv}{HCONV\xspace}
\newcommand{\ngram}{NGRAM\xspace}
\newcommand{\emdh}{EMDH\xspace}
\newcommand{\gate}{GATE\xspace}
\newcommand{\hfreq}{HFREQ\xspace}
\newcommand{\hcomp}{HCOMP\xspace}
\newcommand{\npack}{NPACK\xspace}

\newcommand{\dcomp}{DCOMP\xspace}
\newcommand{\ccheck}{CCHECK\xspace}
\newcommand{\csel}{CSEL\xspace}
\newcommand{\strctrl}{SC\xspace}
\newcommand{\dtw}{DTW\xspace}

\def\BibTeX{{\rm B\kern-.05em{\sc i\kern-.025em b}\kern-.08em
    T\kern-.1667em\lower.7ex\hbox{E}\kern-.125emX}}

\newcommand{\arch}{Hull\xspace}

\title{A Multi-Site Accelerator-Rich Processing Fabric for Scalable Brain-Computer Interfacing}
\author{
    {Karthik Sriram, Raghavendra Pradyumna Pothukuchi, Michał Gerasimiuk,     Oliver Ye, Muhammed Ugur}\\
    {Rajit Manohar, Anurag Khandelwal, Abhishek Bhattacharjee}\\
    {Yale University, New Haven, USA}
}

\begin{document}

\maketitle

\thispagestyle{plain}
\pagestyle{plain}

\begin{abstract}
\arch\footnote{A hull is the protective outer covering of grain. We call our design \arch since it similarly protects the brain.} 
is an accelerator-rich distributed implantable brain-computer interface (BCI) that reads biological neurons at data rates that are 2-3 orders of magnitude higher than the prior art, while supporting many neuroscientific applications. Prior approaches have restricted brain interfacing to tens of megabits per second in order to meet two constraints necessary for effective operation and safe long-term implantation---power dissipation under tens of milliwatts and response latencies in the tens of milliseconds. \arch also adheres to these constraints, but is able to interface with the brain at much higher data rates, thereby enabling, for the first time, BCI-driven research on and clinical treatment of brain-wide behaviors and diseases that require reading and stimulating many brain locations. Central to \arch's power efficiency is its realization as a distributed system of BCI nodes with accelerator-rich compute. \arch balances modular system layering with aggressive cross-layer hardware-software co-design to integrate compute, networking, and storage. The result is a lesson in designing networked distributed systems with hardware accelerators from the ground up.

\end{abstract}

\section{Introduction}

Brain-computer interfaces (BCIs) sense the electrical activity of biological neurons and electrically stimulate them to ``rewire" neuronal circuits. By directly connecting brains to computers, BCIs help advance our understanding of the brain and the mind \cite{andersen_exploring_2022, lebedev_brain-machine_2017}, offer treatment of neurological disorders~\cite{lebedev_brain-machine_2017,chandrasekaran_historical_2021, widge_affective_2014,shih_brain-computer_2012,mcfarland_therapeutic_2017}, enable industrial robotics~\cite{fazel-rezai_brain_2013}, permit novel modes of personal entertainment~\cite{muhl:survey}, and more. 

BCIs can be realized as surface electrodes (i.e., electrical sensors) placed on the scalp above the skull 
to measure brain activity~\cite{chandrasekaran_historical_2021,lebedev_brain-machine_2017}. 
While such wearable BCIs do not require surgical deployment, the signals they collect are muffled by the skull, making them noisy, low-resolution, and less ideal 
for forward-looking BCI applications~\cite{buzsaki:origin, pesaran:investigating, milan:invasive, musk:integrated, shih_brain-computer_2012}. 

Instead, this work focuses on implantable BCIs that are surgically embedded directly on, around, and in the brain tissue~\cite{thakor_distributed_2021, rapeaux_implantable_2021}. 
Implantable BCIs directly record from and stimulate 
neurons with high fidelity, spatial resolution, and in real time~\cite{shih_brain-computer_2012, lebedev_brain-machine_2017}. 
Hundreds of individuals use clinically approved implantable BCIs to treat epilepsy, movement disorders, as well as impaired vision~\cite{fda_affairs_2021, neuropace:rns, medtronic:activapc}. Implantable BCIs are also being studied in 
clinical trials to assess their effectiveness in treating brain stroke, memory disorders, paralysis, anxiety/depression, addiction, and more~\cite{rapeaux_implantable_2021, fcc:new-guidance, jason_synchron_2021}. 

Conflicting constraints make it challenging to design hardware for implantable BCIs. BCIs cannot overheat brain regions by  $>$1$\,^\circ$C to avoid  cellular damage~\cite{serrano-amenos_thermal_2020,wolf:thermal} and must therefore be ultra-low-power. But, BCI designers are also seeking to leverage improvements in sensor technology that are reading exponentially increasing neuronal data \cite{stevenson:advances}. It is challenging to constrain power/energy while processing such large data, especially to respond to neuronal activity in real-time ({\it i.e.,} in ms).
Hardware over-specialization is not a viable way to reduce BCI power; to enable many research and clinical studies, BCIs must be adequately programmable to personalize algorithms, support several computational methods to treat multiple disorders, and enable deployment of maturing/emerging algorithms~\cite{karageorgos:halo, zelmann_closes_2020}.

Complicating BCI design further is the emergence of applications that read and process neural activity from many brain sites over time~\cite{jirsa_virtual_2017,bartolomei_defining_2017,andersen_exploring_2022, thakor_distributed_2021}. This is because the brain's functions (and disorders) are ultimately based on physical and functional connectivity between brain regions that evolve over time~\cite{jirsa_virtual_2017,bartolomei_defining_2017,andersen_exploring_2022, thakor_distributed_2021}.

Existing BCIs~\cite{kassiri:closed, leary:nurip, medtronic:activapc, neuropace:rns, karageorgos:halo} are designed for single-site implantation and lack the ability to store adequately long historical neural data.
Most BCIs~\cite{aziz:256:delta, chen:hardware} have additional limitations in that they
have historically eschewed a subset of programmability, data rates, and flexibility to meet safe power constraints. These BCIs are specialized to a specific task and/or limit personalization of the algorithm~\cite{aziz:256:delta, chen:hardware}. Some support more  programmability by sacrificing high data rates~\cite{medtronic:activapc, neuropace:rns, kassiri:closed, leary:nurip}. 
Consequently, none support the distributed and rapidly evolving neural data processing that emerging BCI applications require. Recent work on HALO balances flexibility, data rates, and power, but is limited to {\it one brain site}. At best, distributed BCI applications have been studied in prior work that consists of multiple sensor implants that offload processing to external devices with higher power budgets~\cite{ahmadi_towards_2019,lee_neural_2021}. But, this is not a panacea because of the long network latency, privacy, and mobility limitations~\cite{zhu_closed-loop_2021}.

Our work is the first to offer a path toward scalable whole-brain interfacing across multiple brain sites. Our solution, \textit{\arch}, is a distributed BCI consisting of multiple BCI nodes 
that communicate with one another wirelessly and interface with the brain with an aggregate data rate 2-3 orders of magnitude higher than the prior state-of-art \cite{karageorgos:halo}. Each BCI node consists of flexible compute made up of reconfigurable power-efficient domain-specific hardware accelerators customized for important neural processing applications. These accelerators are tightly co-designed with storage and networking to ensure system-wide adherence to power and response latency constraints. 

\arch uses a scheduler based on an integer linear program (ILP) to optimally map tasks and algorithms to the hardware accelerators across all of \arch's nodes, 
and to create the network and storage access schedules to feed the accelerators. \arch 
supports three types of BCI applications~\cite{zrenner_closed-loop_2016, translation_challenge}: 

The first category consists of applications that continuously monitor brain activity and respond to aberrant behavior without engaging with agents external to \arch~\cite{hebb_creating_2014,zrenner_closed-loop_2016,translation_challenge}. This includes, for example, detection of seizures~\cite{bartolomei_defining_2017}, prediction of their spread, and finally, to mitigate symptoms, electrical stimulation of regions where seizures are expected to migrate.  

The second category 
also monitor the brain continuously but rely on agents external to \arch to respond. This includes, for example, the detection of an individual's intended movement, and relaying of this data to prostheses or assistive devices~\cite{zhu_closed-loop_2021, chandrasekaran_historical_2021,zrenner_closed-loop_2016,translation_challenge}. These applications enable paralyzed individuals to actuate machines and restore (partial) sensory function.

The third category interactively queries \arch to analyze data from multiple brain sites. These queries may be used ``in the loop" or ``out of the loop". With the former, clinicians may need to modify treatment options based on data that the BCI reads; e.g., confirming that \arch has correctly detected seizures and responded appropriately~\cite{sladky_distributed_2022,kural_accurate_2022}. The latter refers to interactive queries used by technicians to debug system operations, by clinicians to glean the individual's medical history, and by researchers to better understand brain function.


In supporting these types of applications, \arch offers research insights on building end-to-end computer systems centered on hardware accelerators. Specifically, our choice of hardware accelerators follows several important design principles:

First, to maximize power efficiency, as well as simplicity of hardware design, we identify algorithmic kernels within our applications that  accelerate not only their computation but also their networking and storage latency needs.
This is a non-trivial exercise that requires domain-specific knowledge to convert BCI applications into equivalent forms that are amenable to exposing these kernels. We reformulate known computational pipelines for seizure propagation prediction/treatment and movement intent -- two BCI applications that \arch focuses on -- to use hash-based similarity measures that identify the neural signals from different brain sites likeliest to be correlated, before applying heavy-weight correlation measures. The same hashes can drive the design of domain-specific layouts of data in our storage stack. Co-designing our hashes with our storage layout permit several power/latency-efficient data retrievals from our storage layer. The hashes also enable reducing the data communicated in the intra-BCI network between the \arch nodes. 

Second, we design our accelerators to be predictable in latency 
and power for our target data rates. Predictable performance and power characteristics facilitate the optimal design of compute and network schedules---the enabling feature for our ILP scheduler. 
Designing accelerators with predictable performance and power requires care. For some accelerators, whose data generation rate is input-dependent (e.g., data compression), we use theoretically-derived worst-case latency and throughput estimates. Furthermore, we design our accelerators in their own clock domains to enable a range of operating frequencies with well-defined power and performance characteristics.

Third, we design our accelerators to be reconfigurable. This permits the repurposing of our hash accelerators, for example, to act as hash indices for our storage layer, as filters that reduce our networking traffic, and to be tuned differently depending on the target application being supported.

Overall, \arch scales brain-computer interface bandwidth beyond what was previously achievable. \arch is flexible, reconfigurable, and supports real-time distributed neural processing.  
 We use a detailed physical synthesis flow in a 28$\,$nm CMOS process (including tapeouts of partial hardware at 12$\,$nm)  
 coupled with network and storage models to evaluate \arch's power and performance. 
We show that \arch can process  an aggregate of 460$\,$Mbps wireless data from multiple regions in only 10$\,$ms and dissipates no more than 15$\,$mW at any node, confirming its suitability for autonomous use. In interactive mode, it can support 10 queries per second over 6 MB of data over 10 implants. Existing designs support two orders of magnitude lower data and need intrusive wiring. In summary, our specific contributions are:
\begin{enumerate}[leftmargin=4mm]
    \item \arch, the first distributed and wireless BCI system that scales to multi-region neural processing in real time. 
    \item The cross-layer co-design of BCI applications, processing, and storage for scalable and distributed neural processing. \arch is the first to support long-term storage of data, hashing, and database indices to enable distributed signal processing on a single BCI platform.
    \item An evaluation of \arch's flexibility and performance on epileptic seizure propagation and detection of movement intent, in various deployment scenarios, as well as in the support of more arbitrary queries that may be used interactively by clinicians/technicians.
\end{enumerate}


\arch furthers the elevation of hardware accelerators to first-class compute citizens that, like CPUs, can directly engage with networking and storage. This trend will be crucial to future IoT, swarm, and intermittent computing environments that sustain adaptive and complex functionality while meeting strict safe-use (i.e., power, latency, throughput) constraints. 


\section{Background}
\label{back}

\subsection{Brain-Computer Interface Design}
\label{sub:bci}

BCI applications typically perform signal measurement, feature extraction, classification/decision-making, and when applicable, neural feedback/stimulation~\cite{lebedev_brain-machine_2017,shih_brain-computer_2012}. The hardware organization of BCIs reflects these four aspects. Signal measurement is performed by electrodes that read the electrical activity of clusters of biological neurons, and analog to digital converters (ADCs) then digitize this data. State-of-the-art sensors consist of 96-256 electrode arrays per implant. ADCs typically encode measured signal samples with 8-16$\,$bits at a rate of 20-50$\,$K samples per second per electrode. Digitized data is then relayed to compute logic for feature extraction, based on which classification/decision-making proceeds. If needed, the electrode arrays are repurposed, after a digital-to-analog (DAC) conversion step, to electrically stimulate the brain. 

Modern BCIs also use radios to communicate with external agents (e.g., servers, prostheses), presenting an evolution from the surgical cables (which were susceptible to infections and restricted free movement) used in early BCIs~\cite{simeral_home_2021,yin_wireless_2014}. Finally, BCIs are powered with rechargeable batteries and/or inductive power transfer. These components are packaged in hermetically-fused silica or titanium capsules. While the power limit considered safe for permanent implantation varies on the implant's target location and depth, we use 15$\,$mW as a conservative limit~\cite{kim:thermal,karageorgos:halo} for each of \arch's constituent BCI nodes.

\subsection{Neural Processing Applications \& Algorithmic Kernels}
\label{sub:appsBack}

Future BCI applications will collect data across multiple brain sites, and compare histories of stored neural signals across them. Many applications exhibit these needs, including algorithms for neuromuscular rehabilitation and neuropsychiatric disorders~\cite{thakor_distributed_2021,lebedev_brain-machine_2017, sporns_graph_2018, bilge_deep_2018, deco_great_2014,widge_affective_2014}, but we focus on epileptic seizure propagation and detection of movement intent as they form the bulk of emerging BCI use~\cite{gallego_going_2022, shih_brain-computer_2012, camargo-vargas_brain-computer_2021,lebedev_brain-machine_2017,bensmaia_restoring_2014}. 
In addition, we also consider spike sorting, a crucial kernel widely used in many applications~\cite{todorova_sort_2014,rey_past_2015}. Spike sorting differs from seizure propagation and movement intent in that it 
is not a full application in itself. Nevertheless, we study it because it is a prime candidate for wide use in a distributed manner. 



\subsubsection{Epileptic seizure propagation application} Seizures often migrate across brain regions~\cite{bartolomei_defining_2017}. Predicting seizure spread can help explain seizure dynamics and offer treatment options. When a seizure is detected at a brain site, seizure propagation algorithms compare neural signals from the originating site against current and past signals collected from other brain sites of interest. Correlation measures are used to detect whether there is a seizure match across brain sites;  i.e., whether a seizure is likely to propagate to another brain region.

\begin{figure}[h]
\centering
\subfloat[Seizure propagation analysis.]{
\centering
    \includegraphics[width=0.80\linewidth]{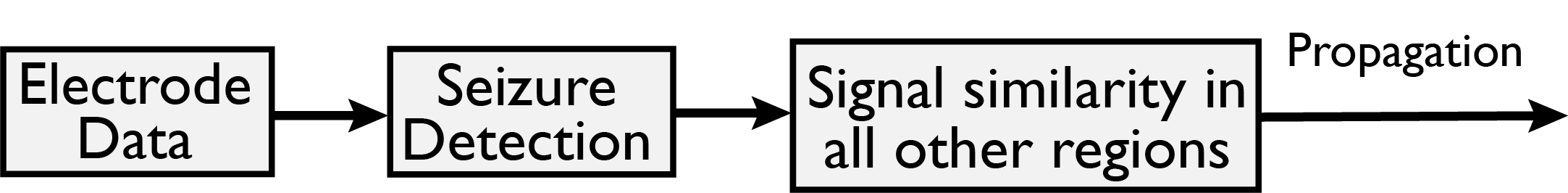}
    \label{fig:seiz_det_background}
}\\ \vspace{-2mm}
\subfloat[Decoding movement intent and stimulating response to it.]{
    \includegraphics[width=0.8\linewidth]{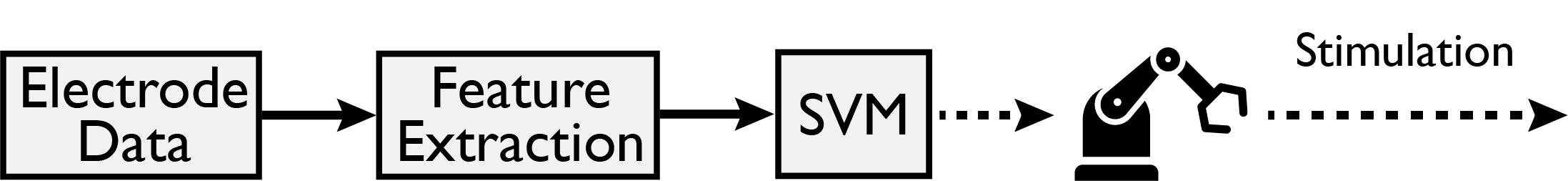}
\label{fig:mov_int_background}
}\\ \vspace{-2mm}
\subfloat[Spike sorting to separate the combined electrode activity.]{
    \includegraphics[width=0.8\linewidth]{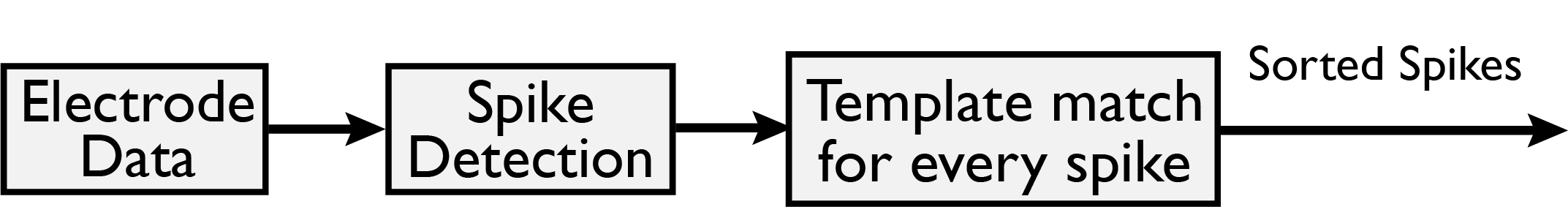}
\label{fig:spike_sort_background}
}\vspace{-1mm}
\caption{Main BCI application steps. BCIs do not yet support on-device seizure propagation or multi-site movement intent.}
\label{fig:apps_background}
\vspace{-6mm}
\end{figure}

Figure~\ref{fig:seiz_det_background} shows the steps used (but unsupported in their entirety in any existing BCI) in standard seizure propagation pipelines~\cite{bartolomei_defining_2017, jirsa_virtual_2017}. First, seizure signals are detected in the signals from each electrode in all the brain regions that the electrodes probe. This step typically uses band-pass filters or a fast Fourier transform (\fft) on continuous signal windows to generate features, followed by a classifier like a support vector machine (\svm)~\cite{shiao:svm}. Alternatively, clinicians may manually annotate the onset of a seizure. 

Once a seizure is detected in a region at a specific point in time, the signal window from that region is compared with all the concurrent and previous windows from all other regions, up to a chosen time in the past.  

\subsubsection{Detection of movement intent application} 
BCIs can infer coarse-grained movement from reading single sites of the motor cortex region~\cite{niazi_detection_2011,jensen_detection_2014}, but more fine-grained movement intent (e.g., the movement of individual fingers grasping an object) requires reading neural activity from multiple brain regions~\cite{gallego_going_2022, ojakangas_decoding_2006,sternad_intention_2009}. Figure~\ref{fig:mov_int_background} shows a typical computational pipeline that infers fine-grained movement intent~\cite{shakeel_review_2015, taylor:direct, fetz_restoring_2015,bensmaia_restoring_2014}. Neural signals from all electrodes in all target brain sites are first filtered or converted into the frequency domain using \fft for feature extraction. Then, the features 
are all pushed into a classifier to deduce intended movement. Linear \svms are commonly used for classification because they are effective, and because their parameters are intuitive for neuroscientists to reason about~\cite{shakeel_review_2015, capogrosso_brainspine_2016, chandrasekaran_historical_2021, muller_linear_2003}. Intended movement is then relayed to an external agent like a prosthetic arm. The prosthetic arm's movement then has to be conveyed to the brain regions (e.g.,  the sensorimotor cortex) responsible for sensing the individual's environment using neural stimulation patterns~\cite{darie_delivering_2017, tafazoli_learning_2020}.

\subsubsection{Spike sorting algorithmic kernel}
Spike sorting is an exemplar of key signal transformations that comprise important applications, and that benefit from engagement with multiple brain sites. Most sensor arrays used in existing BCIs have electrodes that measure the combined electrical activity of a cluster of neurons, rather than that of individual neurons. Spike sorting detects all the peaks in the combined electrode activity and separates them into a series of constituent signal spikes from distinct neurons. Figure~\ref{fig:spike_sort_background} shows this algorithm. It measures the distance of each signal peak from several spike templates, and the nearest template is chosen as the peak's spike. In some variants~\cite{rutishauser_online_2006}, the templates are obtained dynamically from clustering the peaks. Spike distances are measured with dynamic time warping (\dtw) or earth movers distance (EMD)~\cite{sotomayor-gomez_spikeship_2020, grossberger_unsupervised_2018}, which are computationally expensive. Modern spike sorting methods are too slow to be deployed online; distributed spike sorting has even higher overheads.

No existing BCIs support the signal processing needed for historical analysis of seizure and movement intent activity emanating from multiple brain sites, and for distributed spike sorting. Most designs use a single implanted device that senses and processes information from the brain region probed by the implant~\cite{kassiri:closed, leary:nurip, medtronic:activapc, neuropace:rns, karageorgos:halo}. 
Some designs use distributed sensors that do not directly connect to computational support~\cite{ahmadi_towards_2019,lee_neural_2021}, and offload data to an external device. But, the lack of on-device distributed processing precludes BCI support for applications that require ms-scale decisions, such as preempting propagation of seizures, or control of prosthetics.

\begin{figure*}[t]
\centering
\subfloat[\arch overview.]{
\includegraphics[width=0.65\textwidth]{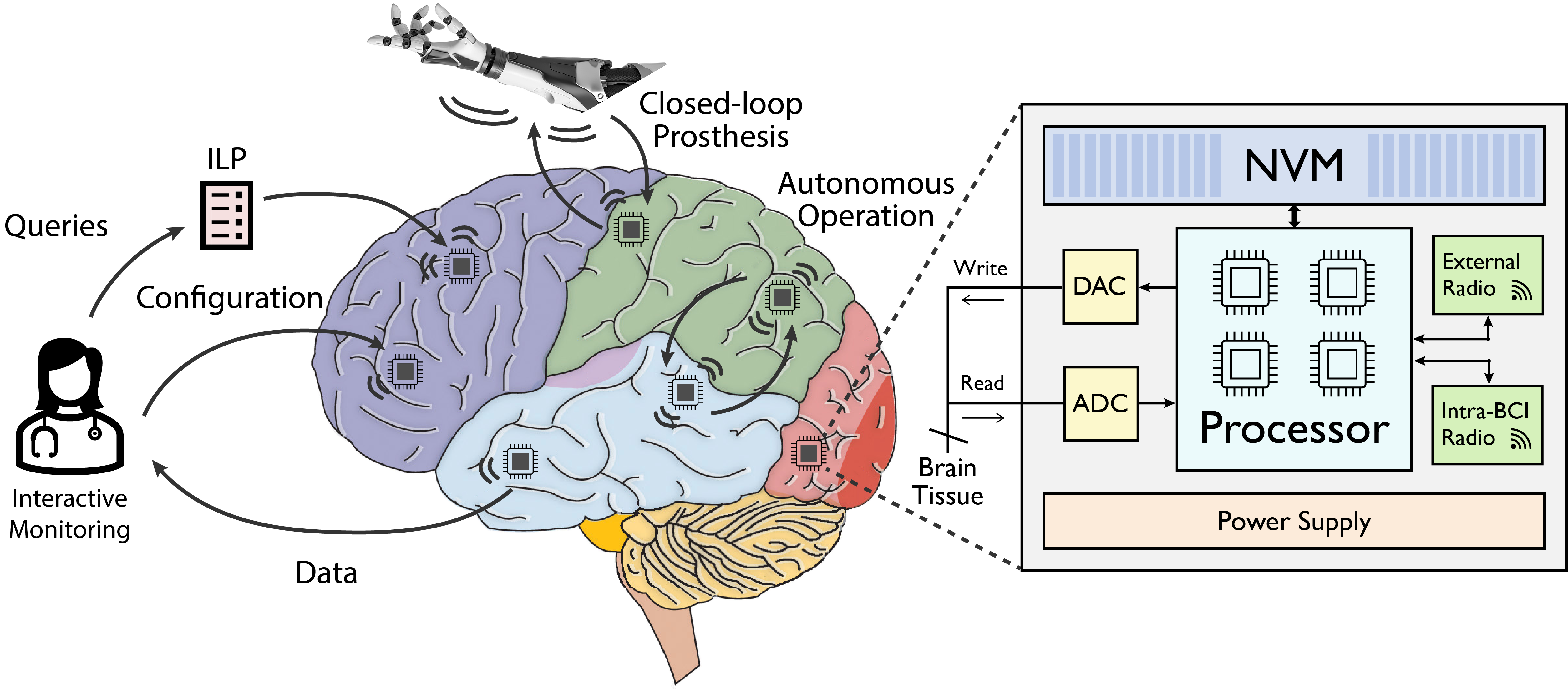}
\label{subfig_arch}
}
\subfloat[The processor fabric in each of \arch's nodes.]{
\includegraphics[width=0.35\textwidth]{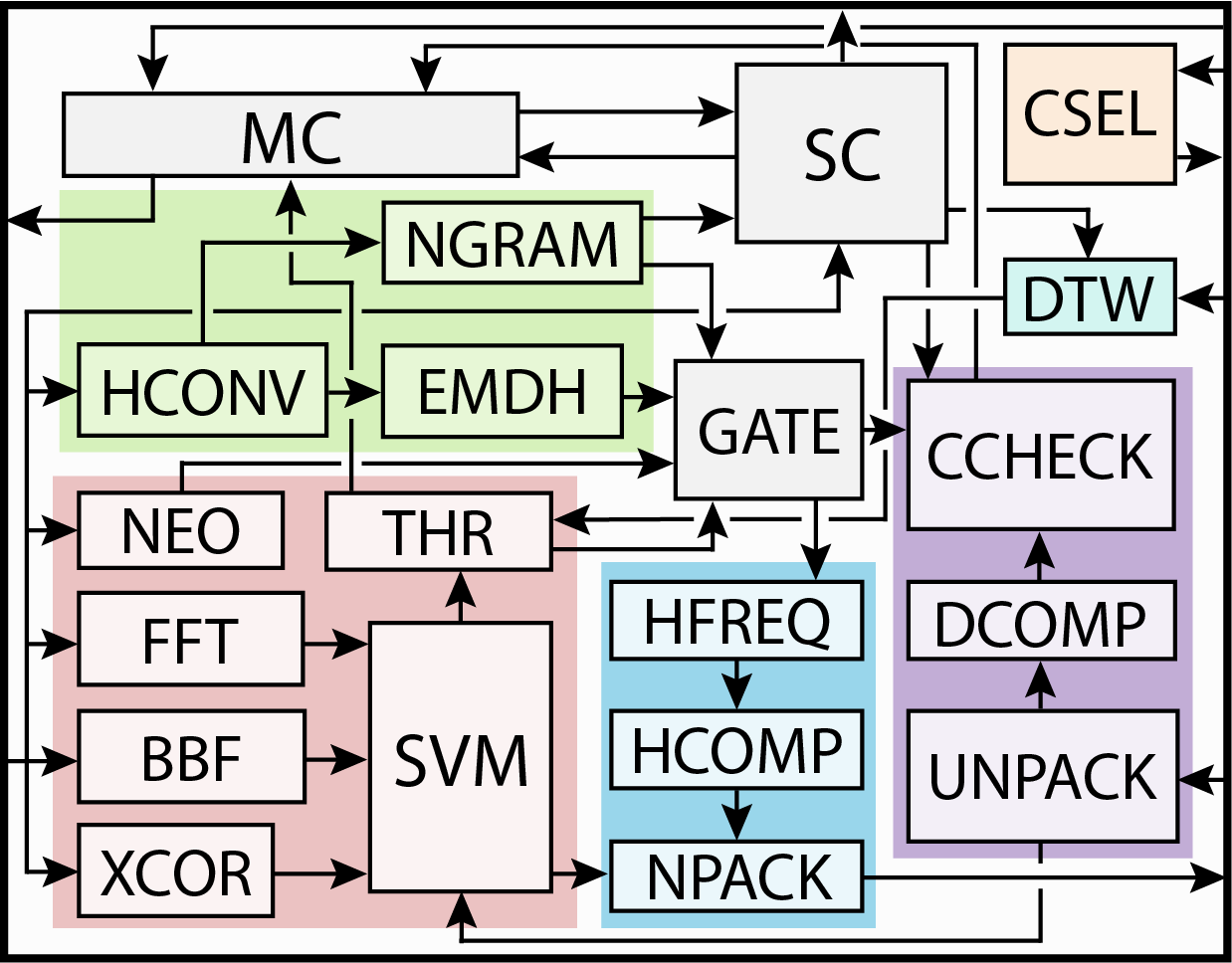}
\label{subfig_proc}
}
\caption{The \arch BCI is made up of nodes that are implanted in distinct brain sites. The nodes communicate wirelessly with each other and external agents. Each \arch node has sensors, radios, analog/digital conversion, processing fabric, and storage; the processing fabric contains hardware accelerators and configurable switches that can be used to create different pipelines.}
\label{fig_overview}
\vspace{-6mm}
\end{figure*}

\subsection{Locality-Sensitive Hashing for Signal Comparison}
\label{sub_hash}

All the applications described previously use signal comparison that is expensive. We use locality-sensitive hashing for fast time series matching~\cite{Kim2017PhysiologicalTS} to meet \arch's ms-scale latency constraints. We face two challenges in using locality-sensitive hashing. The first is the presence of variable-latency computations involving randomization, and the other is the need to support multiple comparison measures---the choice of measure varies across BCI uses~\cite{cao_real-time_2016,sotomayor-gomez_spikeship_2020, grossberger_unsupervised_2018}. We leverage prior work on two locality-sensitive hashing schemes developed for \dtw~\cite{luo_ssh_2016} and EMD~\cite{gorisse_locality-sensitive_2012}. Subsequent sections describe how we modify them to suit the needs of \arch's target applications. 

The \dtw hash generation process~\cite{luo_ssh_2016} first creates sketches of the signal by using the dot product of a random vector with sliding windows in the signal. If the dot product is positive, the sketch value for the window is 1; otherwise, it is 0. Next, it counts the occurrences of n-grams formed by $n$ consecutive sketch values. The n-grams and their counts are used by a randomized weighted min-hash to produce the final hash. 

The original EMD hash~\cite{gorisse_locality-sensitive_2012} is obtained by first calculating the dot product of the entire signal with a random vector, and computing a linear function of the dot product's square root.

\subsection{Flexibility as a Goal in Brain-Computer Interface Design}
\label{environments}

A key takeaway from Sections \ref{sub:appsBack} and \ref{sub_hash} is the need for flexible support of compute on emerging BCIs. Indeed, this is a topic explored in recent work on the HALO architecture for BCIs \cite{karageorgos:halo, bhattacharjee_halo_2022, karageorgos_balancing_2021}. Prior to HALO, power efficiency was achieved by specializing BCIs to offer a specific type of computation for a specific brain region. However, flexibility is an important requirement for future BCIs for several reasons:

First, there is no single {\it best} signal processing pipeline for a task; instead, there exist several distinct signal processing pipelines with different tradeoffs~\cite{zelmann_closes_2020,translation_challenge,sun_closed-loop_2014}. For \arch, this means that the specific hardware accelerators needed to support target computational pipelines (e.g., \dtw vs cross-correlation), and the configuration of key parameters in these accelerators (e.g., window sizes, thresholds) must be customizable to users.

Second, BCIs may be used in different ways~\cite{zrenner_closed-loop_2016,translation_challenge}. One use is autonomous operation, monitoring neural activity and stimulating neurons when a harmful event occurs. An example is epileptic seizure monitoring and deep brain stimulation to preempt the seizure before its onset~\cite{sun_closed-loop_2014}. Alternatively, BCIs may translate neural activity into commands for an external device~\cite{zhu_closed-loop_2021} (e.g., the commands to move a prosthetic) or the letters to be displayed on a screen~\cite{shih_brain-computer_2012}. It is common for the BCI to also translate the external activity into neural feedback (e.g., to recreate the sense of touch and movement)~\cite{vidal_review_2016}. 

Third, beyond clinical uses, the same BCI platform should support algorithmic deployment and data collection for research and exploration of the brain sciences~\cite{shih_brain-computer_2012,translation_challenge,herff_potential_2020}. In these cases, many applications and usage modes may be necessary depending on the desired experiment.
Some of these uses may require interactive monitoring, where the BCI and a clinician are part of the decision-making loop~\cite{sladky_distributed_2022}. In this case, the BCI operates autonomously until it detects abnormal activity, such as the onset of a seizure. When this happens, it alerts a clinician, who can use additional data from the individual to determine the course of action~\cite{sladky_distributed_2022}. A useful BCI system must be customizable to support these different scenarios.

Beyond these scenarios, there are many practical reasons that BCIs should be flexible, such as changes in the  individual's neurological conditions (which may require modifying treatment protocols), changes in electrode behavior from the immune response of the brain to the BCI etc.~\cite{translation_challenge, shih_brain-computer_2012, sun_closed-loop_2014}.

Supporting high performance with flexibility under extreme power and latency constraints is challenging. Like HALO, \arch relies on modular hardware accelerators (henceforth referred to as processing elements or PEs) to form various signal processing pipelines. Unlike HALO and any existing BCI, however, \arch supports the distributed signal processing applications in Section \ref{sub:appsBack} for the first time.

 \section{The Design of the Hull System}
\label{arch}

Figure~\ref{fig_overview} shows the \arch BCI and its constituent \arch nodes implanted in different regions of the brain. \arch nodes communicate with one another wirelessly. An ILP scheduler maps applications and interactive queries onto \arch's nodes. 

Each \arch node contains 16-bit ADCs/DACs, a reconfigurable processor with several PEs, an integrated physical storage layer made of non-volatile memory (NVM), separate radios for \arch's nodes to communicate with one another (i.e., intra-BCI radios) and externally (i.e., external radio), and a power supply. 

\subsection{Rewriting Applications for On-Device Processing}
\label{sub:apps}

We make three changes to existing BCI applications to run them on \arch (to meet real-time constraints), rather than relying on external processing. First, we rewrite the signal processing pipelines to use fast hash-based signal comparison in the common case, falling back to more time-consuming approaches (e.g., cross-correlation or \dtw) only when more accurate computation is really necessary. Second, we allow our applications to use memory. Third, we observe that classifiers commonly used in neuroscience are linear (e.g., \svms), and therefore compute classifier outputs hierarchically across \arch's nodes in a manner that reduces network communication.

\begin{figure}[h]
\centering
\subfloat[Seizure propagation.]{
\centering
    \includegraphics[width=\linewidth]{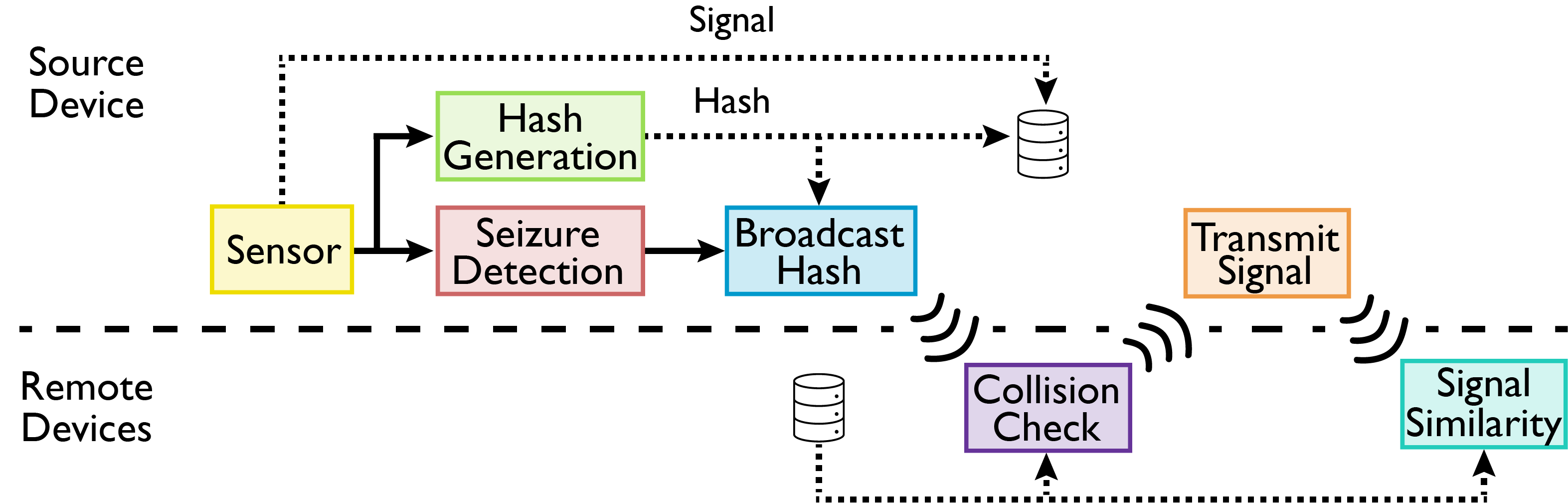}
    \label{fig:seiz_det_high_level}
}\\ \vspace{-2mm}
\subfloat[Decoding movement intent and stimulating response to it.]{
    \includegraphics[width=\linewidth]{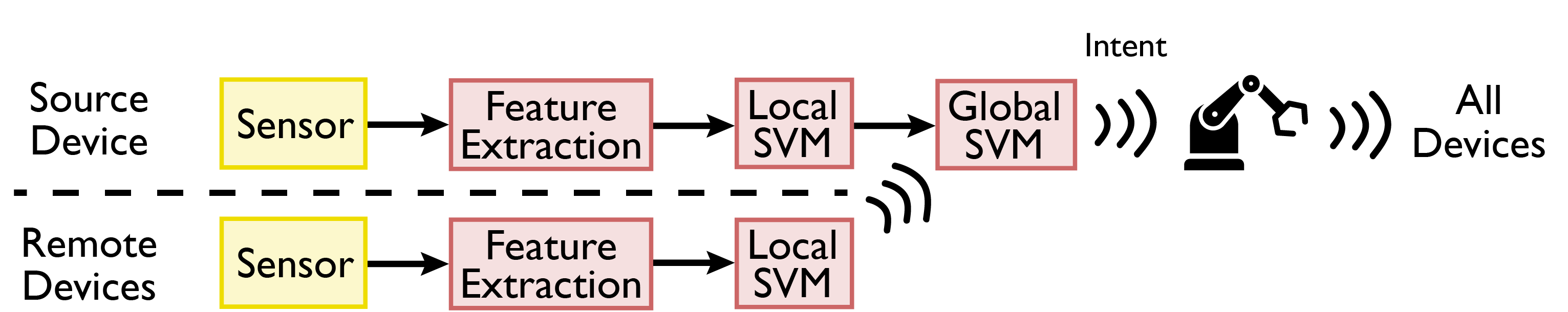}
\label{fig:mov_int_high_level}
}\\ \vspace{-2mm}
\subfloat[Spike sorting.]{
\includegraphics[width=0.8\linewidth]{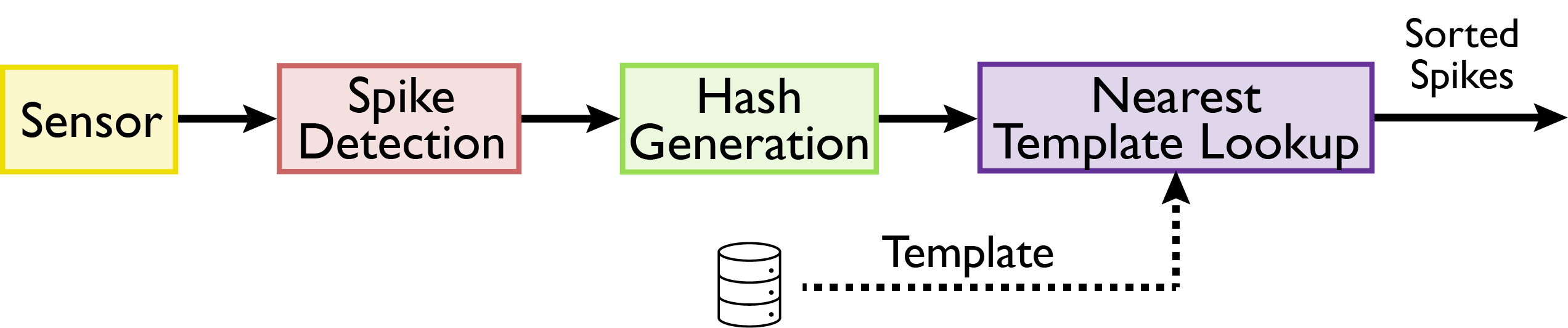}
\label{fig:spike_sort_high_level}
}\vspace{-2mm}
\caption{High-level overview of the BCI applications supported for online distributed processing in \arch.}
\label{fig:apps}
\vspace{-6mm}
\end{figure}

Figure~\ref{fig:seiz_det_high_level} shows our newly created seizure propagation application. While functionally equivalent to the standard version, our application is made up of three phases---seizure detection, hash comparison, and exact signal matching.  On every sample at all electrodes, we generate new hashes for each sliding signal window  (e.g., one hash for a 120-sample window), and store them on the on-device non-volatile memory in each \arch node (Section~\ref{sub_arch}). When a \arch node detects a seizure locally (i.e., in the brain region that it probes), it broadcasts the hashes of the signal windows that were classified as a seizure. All other \arch nodes check if these hashes match with any of their recently stored local hashes, and respond when a match is found. A match indicates that a seizure experienced in one brain region likely has a correlated seizure in another region. To ascertain this, the \arch node that initially detected the seizure broadcasts the entire signal window for the signals that resulted in a hash collision. Seizure propagation is then confirmed by running an exact comparison with these signals at the nodes that had the hash collision. Since the full signal data and exact similarity matches are performed only when necessary, computation per \arch node and communication among \arch nodes is reduced by two orders of magnitude compared to the baseline application pipelines in Section \ref{sub:appsBack}.

Figure~\ref{fig:spike_sort_high_level} shows that we use a similar approach to enable, for the first time, an online version of spike sorting even in distributed scenarios. Like seizure propagation, spike sorting benefits from hash-based signal processing and memory. The templates are stored in NVM, and distance computation is replaced with hash collision checks. Because spike sorting is a precursor to many neural processing algorithms~\cite{todorova_sort_2014,translation_challenge}, this online realization of it for the first time unlocks the ability to support many spike sorting-centered applications.

Finally, movement intent also benefits from computing our linear classifier hierarchically. Figure~\ref{fig:mov_int_high_level} shows the pipeline that \arch supports. Each \arch node computes a partial classifier output from the signals it receives and transmits the output. One node, the leader, computes the final \svm classification and communicates it to an external prosthetic device. The prosthetic device's movements are broadcast back to \arch; each node then electrically stimulates the sensorimotor cortex of the brain to simulate the ``feeling" of having moved a natural limb.

\subsection{Flexible \& Energy-Efficient Accelerator Design}
\label{sub_arch}

Figure~\ref{subfig_proc} shows the processing fabric that we design for each of \arch's nodes. Several accelerators or PEs are connected via programmable switches to realize many signal processing pipelines. A low-power microcontroller (MC) support miscellaneous workloads for which there are no PEs. The PEs are designed for flexibility to support various computational functions, power/energy- and area-efficient acceleration, and deterministic latency and energy consumption to enable our ILP scheduler to optimally map application tasks onto our accelerators.  We use the recently-published HALO architecture~\cite{karageorgos:halo} as a starting point to realize a set of PEs that are useful for single-implant scenarios, and then go beyond to realize PEs that accelerate our distributed neural applications. 

\arch includes PEs for single-site spike detection (NEO--non-linear energy operator; DWT--discrete wavelet transform), compression (LZ4; LZMA), feature extraction (\fft--discrete fast Fourier transform; \xcor--cross-correlation measure; \bbf--Butterworth bandpass filtering), thresholding (\thr), conditional (\gate), classification (\svm--linear support vector machine), and the radio for communication with systems outside of \arch. 

\arch then integrates several new PEs to support distributed computation, fine-grained wireless communication, and access to per-node NVM. Each PE has appropriately sized SRAM buffers to support its processing. The PEs include support for:


\subsubsection{Hash generation} \arch supports Euclidean, cross-correlation, \dtw, and EMD; we support configurability of hash settings for all four measures. 

First, we identify that important parameters of the \dtw hash (e.g., size and step of the sliding window), and n-gram length (Section~\ref{sub_hash}) can be modified to also support Euclidean, and cross-correlation measures. There is no need for new hardware to support additional means of configurability beyond what is already needed for the \dtw-hash parameters. 

Second, we identify that the \dtw and EMD hashes share  dot product computation of the signal with a random vector (Section~\ref{sub_hash}), enabling the reuse of hardware.

Finally, we select a different weighted min-hash algorithm for the last step of the \dtw hash than the one originally proposed in prior work~\cite{luo_ssh_2016}. Our approach~\cite{consistenthash} preserves hash properties while achieving deterministic latency and power. 

Our hash generation uses three PEs: \hconv, to obtain the dot product of a configurable signal window with a random vector; \ngram, to compute the n-gram counts in a signal and generate the \dtw-based hash; and \emdh, to square root the dot product, and other operations to generate the EMD hash. 

\subsubsection{Hash collision check} To determine signal similarity across multiple brain sites, the hashes received over the network by the \arch nodes must be compared with the locally generated hashes in the recent past (e.g., 100$\,$ms). Each \arch node uses a \ccheck PE that receives decompressed hashes from the network, stores them in SRAM registers, and sorts them in place. The PE requests the storage controller (\strctrl) to read the hashes to be compared from the NVM. 
These hashes are compared with those in the registers using binary search. 



\subsubsection{Signal similarity} \csel identifies signals for exact signal comparison using \dtw, EMD, and Euclidean distance. For \dtw, we build a pipelined implementation that uses the standard \dtw algorithm~\cite{dtw} with a Sakoe-Chiba band parameter for faster computation~\cite{sakoe:dynamic}. 
This PE can also support Euclidean distance computation by using the Sakoe-Chiba band parameter to be 1. We use the microcontroller to run EMD~\cite{pele:fastemd} for now, although we will build custom PEs in the future.





\subsubsection{Intra-BCI network compression and packing} The intra-\arch network transmits hashes and signals. We compress the hashes but transmit uncompressed raw signals. Compression makes data more vulnerable to bit errors. Because the hashes are used only for approximate matching, bit errors are not as critical to the quality of signal correlation. But, the raw signals are used for accurate matching. Measures like \dtw are naturally resilient to single-bit errors in the signal, but their quality worsens rapidly with erroneous compressed signals. 

Compression PEs (i.e., LZ/LZMA) built for HALO do not meet \arch's power and latency constraints for hashes. Instead, we build PEs customized to our particular data/communication needs. The \hfreq PE collects the hash values (and sorts them by frequency of occurrence) that a \arch node must transmit. The \hcomp PE encodes the hashes first with dictionary coding, then uses run-length encoding of the dictionary indexes~\cite{pu_fundamental_2005}, and finally uses Elias-$\gamma$ coding~\cite{elias_code} on the run-length counts. \hcomp's compression ratio is only 10\% lower than that of LZ4/LZMA, but consumes $\approx$7$\times$ less power.


Compressed data is sent to the \npack PE, which adds checksums before transmission. The UNPACK and \dcomp PEs decode and decompress packets on the receiving side. 

\subsubsection{Storage control} An SC PE manages NVM access. SC uses SRAM to buffer data before NVM writes in 4$\,$KB pages. The SRAM also permits data buffering during NVM erase operations when writes cannot be accepted. Finally, SC (and the SRAM) permits data reorganization to accelerate future reads from the NVM (Section~\ref{subsub_nvm}). \strctrl uses registers to store metadata about data written by the ADC and hash PEs (e.g., the last written page and the size of written data). This accelerates, for example, the search for recent common signal data.





\subsubsection{Microcontroller} The MC runs at low frequency (20$\,$MHz), and integrates 8$\,$KB memory. It configures individual PEs into target pipelines (Section~\ref{implement}) and receives commands to stimulate neurons either for stopping a seizure or for conveying neural feedback from a prosthetic. The MC can be used for general-purpose computation not supported by any PEs such as new algorithms, or infrequently run system operations such as clock synchronization (Section~\ref{sub_clock}).

\subsubsection{Well-defined throughput} Each PE operates in its own clock domain, like prior work \cite{karageorgos:halo}, but also supports multiple frequencies. This enables each PE to lower operating frequency (and reduce power) to the minimum necessary to sustain the PE's target data rate, for varying input electrode counts. This feature also ensures fixed latency even when PEs process a variable number of input electrode signals. We design each PE to support a maximum frequency  $f_{max}^{PE}$ which is high enough to support the maximum data processing rate required. We use a configurable register that can be used to set the frequency to $f_{max}^{PE}/k$, where $k$ is user-programmable. The clock frequency is varied using a simple state machine that uses a counter to only pass through every $k$ clock pulses. The power consumed by this counter is in the $\mu$W range~\cite{bingham2018qdi}, much lower than the per-PE power. Overall, the dynamic power of the PEs scales linearly with the frequency. This also enables deterministic power and latency and helps optimal scheduling (Section~\ref{sub_sched}). 


\subsection{On-device non-volatile memory}
\label{subsub_nvm}

Each \arch node integrates 128$\,$GB on-device NVM to store raw neural signals, hashes of these signals, and pre-loaded data needed by applications (e.g., templates for spike sorting). We divide the NVM into four partitions, one for each of these classes of data, and another for use by the MC. The sizes of the partitions are configurable through registers. When a partition is full, the oldest data in the partition is overwritten.

We optimize the layout of signal and hash data  in the NVM for performance and power. \arch's ADCs (and hash generation PEs) process electrode samples sequentially. If the data is stored in this manner, extracting a contiguous segment of one signal would require long-latency reads from multiple discontinuous NVM locations. Instead, we store contiguous chunks (where a chunk size is user-specified) of each signal. Retrieving the signal (or hashes) at a particular electrode and time-step need only offset calculations. SC enables this reorganization as it buffers data in 4$\,$KB SRAM pages before NVM writes.

\begin{figure*}[h]
    \centering
    \includegraphics[width=0.8\textwidth]{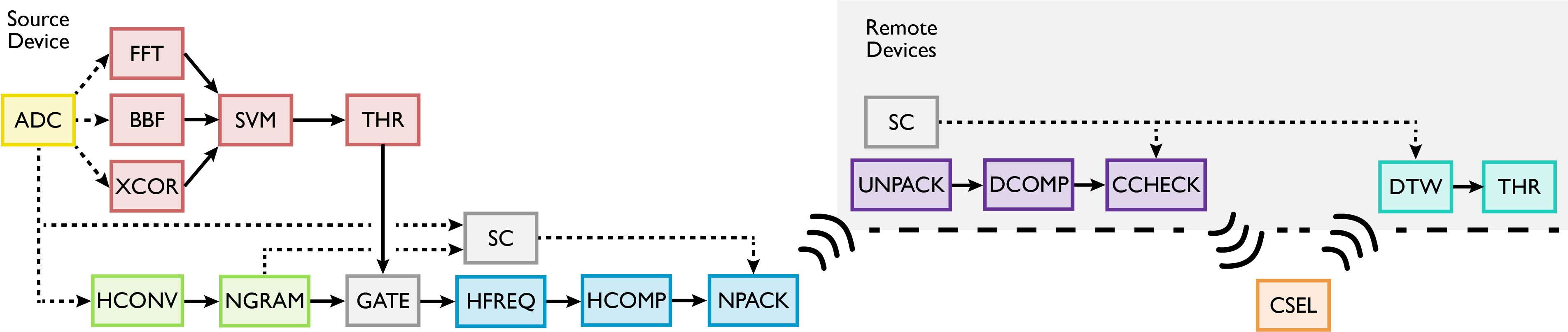}
    \caption{Seizure detection and propagation on \arch. The colors of the PEs are matched with the high-level tasks from Figure~\ref{fig:seiz_det_high_level}. }
    \vspace{-6mm}
    \label{fig:seizure_detection_pipeline}
\end{figure*}

\subsection{Networking}
\label{sub_network}



We use separate radios for intra-\arch and external device communication as the required distances and communication needs are different. For intra-\arch communication, we use a custom network protocol with a fixed schedule across the nodes. The schedule is decided by an ILP based on application goals (Section~\ref{sub_sched}). To coordinate intra-\arch communication, we use TDMA for its simplicity and deterministic behavior. 


Each network packet has an 84-bit header, and a variable data size up to a maximum packet size of 256 bytes. The header and the data have 32-bit CRC32~\cite{crc} checksums.
On a checksum mismatch, the receiver simply discards the packet and does not participate in the pipeline for processing the current sample. However, we find that while it is best to discard erroneous packets with hashes, erroneous packets carrying raw signal data can still be used without adversely affecting the overall application because of the resiliency of measures like DTW. 


\subsection{Task Scheduling on Accelerators, Storage, \& Networking}
\label{sub_sched}

As input to our ILP scheduler, users provide a description of the desired computation as a dataflow pipeline using functions of the PEs, or as an interactive query from which the dataflow can be extracted (Section~\ref{implement}). They also provide the priorities of the tasks in the application (e.g., seizure detection versus signal comparison), and constraints like the overall response latency. A higher priority task ensures that the system processes more neural signals in this task relative to the others when all signals cannot be processed for all tasks due to power or latency constraints.
 The ILP maps each function to the corresponding PE in one or more of \arch's nodes.


The ILP considers each possible mapping of application tasks (e.g., seizure detection, hash comparison) to PEs as a flow, and maximizes the weighted sum of the number of channels processed in each flow. It uses three major constraints:

\noindent
\textit{Latency:} End-to-end latencies through the PEs and communication links must be below a specified limit.

\noindent
\textit{Power: }The power consumed by all the PEs and links at all times must be below a specified limit.

\noindent
\textit{Communication: }Only one flow is allowed to use the radio at any time because of TDMA. 

Our ILP setup is simple because of the behavior of the PEs. With variable throughput processing, the latency of processing any number of input signals is the same. The dynamic power consumed by a PE scales predictably linearly with the input size (since frequency scales linearly). Finally, the system allows two flows to share the same PE. When this occurs and electrode signals to be processed are allocated to both flows, the signals from each flow are interleaved so that they are all run at the same frequency---completing within the same time as if they were run independently. The hardware tags the signals from each flow so that they are routed to the correct destinations.


  \subsection{Clock Synchronization} 
\label{sub_clock}

\arch's distributed processing requires the clocks in each BCI node to be synchronized up to a few $\mu$s of precision. \arch's clocks are based on pausable clock generators and clock control units~\cite{yun:pausible,moreno:synthesis} that suffer only picoseconds of clock uncertainty, a scale much smaller than our $\mu$s target. \arch operates at the temperature of the human body and does not experience clock drift due to temperature variance. Nevertheless, \arch runs clock synchronization once a day using SNTP~\cite{mills_1995}.

One of the \arch nodes is set up to act as the SNTP server, to which all other nodes send messages to synchronize their time. The clients send their previously synchronized clock times, and current times, while the server sends its corresponding times. The difference between these values is used to adjust the clocks. This process repeats a few times until all the clocks are synchronized within the desired precision. During clock synchronization, the intra-\arch network is unavailable for application use. However, operations that do not require the network (e.g., seizure detection) or NVM access can continue.

\section{Deploying \arch for BCI Applications}
\label{implement}

\arch supports autonomous epileptic seizure propagation in autonomous, movement intent detection for closed-loop prosthesis, online spike sorting, and interactive querying.  

\vspace{1mm}
\noindent
\textbf{Autonomous seizure propagation and detection: }
Figure~\ref{fig:seizure_detection_pipeline} shows \arch's implementation of autonomous seizure detection and propagation. The choice of the PE functions is based on prior work~\cite{shiao:svm}. This implementation uses \xcor, \bbf, and \fft to extract features from the ADC measurements and uses an \svm to detect a seizure. When a seizure is detected, the nodes exchange hashes for comparison. To confirm that a seizure is indeed likely being propagated, \arch uses the \dtw distance of the signals across nodes, and electrically stimulates the brain in response to predicted propagation within 10$\,$ms.

The dataflow in Figure~\ref{fig:seizure_detection_pipeline} is fed to the ILP to schedule this application on \arch. The ILP generates an optimal mapping of the functions and generates a configuration code. This code is run by each \arch node's microcontroller to configure the PEs.

\begin{figure}[h]
\vspace{-1mm}
\centering
\includegraphics[width=0.6\linewidth]{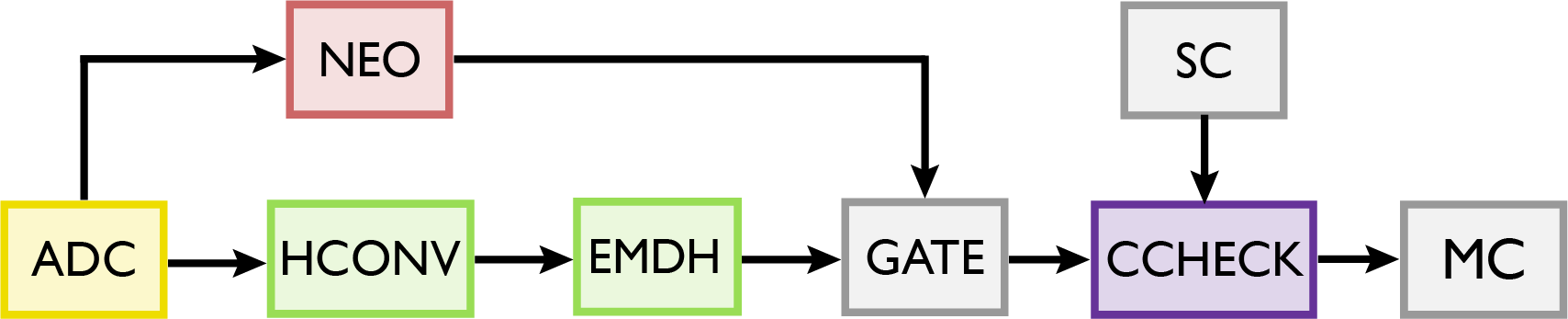}
\caption{Spike sorting on \arch.}
\label{fig:spike_sort_pipeline}
\vspace{-4mm}
\end{figure}

\vspace{1mm}
\noindent
\textbf{Online spike sorting: } Figure~\ref{fig:spike_sort_pipeline} shows the mapping of online spike sorting to \arch. The template-matching version pre-loads the NVM in the nodes with templates and their hashes.

\begin{figure}[h]
    \centering
    \includegraphics[width=\linewidth]{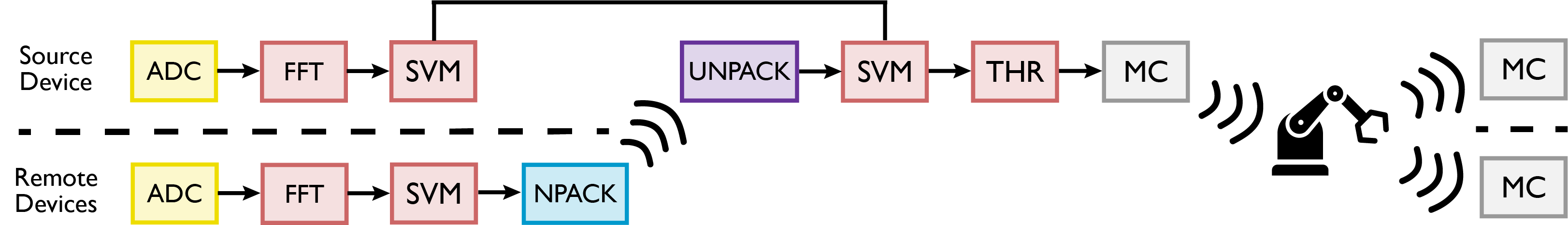}
    \caption{Query pipeline for movement intent application}
    \vspace{-3mm}
    \label{fig:movement_intent_pipeline}
\end{figure}

\vspace{1mm}
\noindent
\textbf{Movement intent detection and feedback: }
Figure~\ref{fig:movement_intent_pipeline} shows how \arch implements detection of movement intent and feedback, augmented from prior work\cite{shakeel_review_2015}. Each node extracts features from its local signals and computes a partial \svm output. Then, one node receives the partial \svm outputs and computes the commands for the prosthetic. The movements of the prosthetic are transmitted wirelessly, and each node runs a stimulation algorithm for its region to provide neural feedback.

\noindent
\textbf{Interactive querying: }
Interactive queries are used to read multi-site data or modify system configuration. The general format for an interactive query follows a select-project structure, akin to SQL queries~\cite{sql}: 

\vspace{1mm}
\noindent
{\small
\texttt{\textcolor{purple}{\textbf{from}} $[$\textit{set of devices}$]$ \textcolor{purple}{\textbf{select}} data[electrodes][time range] \textcolor{purple}{\textbf{where}} condition}
}
\vspace{1mm}

The query specifies \emph{select} criteria, i.e., the range of time from which data is requested, along with the nodes from which the data should be returned, and the \emph{project} criteria, i.e., the conditions that the selected data must satisfy. Similar to select-project-based SQL queries, \arch's interactive query interface can support a wide range of complex queries. The project conditions are evaluated on the PEs when possible, and on the microcontroller otherwise. The following illustrates an example query to fetch $\pm$100$\,$ms of data from all devices from the time they detected a seizure in the last 5$\,$s. This example requires seizure detection using 120$\,$ms windows of the raw signal data.

\vspace{1mm}
\noindent
{\small
\texttt{\textcolor{purple}{\textbf{from}} \textit{*} \textcolor{purple}{\textbf{select}} data[:][t-100:t+100] \textcolor{purple}{\textbf{where}} \textcolor{blue}{\textbf{seizure\_detect}}(data[t-120:t]) \textcolor{purple}{\textbf{and}} t $>=$ -5000 \textcolor{purple}{\textbf{and}} t $<=$ 0  }
}
\vspace{1mm}

Complex examples can supply template signals and request data from nodes that recorded signals similar to the templates. Queries are separately compiled and the extracted dataflow is sent to the ILP, which finalizes query execution schedules. 

Users can also set up the pipelines of specific tasks; e.g., a clinician may modify \texttt{\textcolor{blue}{\textbf{seizure\_detect}}} to use only \fft for feature extraction instead of \fft, \bbf and \xcor as in Figure~\ref{fig:seizure_detection_pipeline}. Such a configuration does not need the ILP.

Interactive queries use a power-hungry radio, precluding simultaneous execution of queries and autonomous tasks in some cases. Some of these are either slowed down or temporarily paused; e.g., when a clinician responds to a seizure alert and requests recent signal data, seizure propagation has to be paused to send the data to the clinician.




\section{Methodology}
\label{setup}

\noindent{\bf Processing fabric: } \arch's PEs are designed with a commercial 28$\,$nm fully-depleted silicon-on-insulator (FD-SOI) CMOS process and synthesized using the Cadence\textsuperscript{\textregistered} suite of tools. We use standard cell libraries from STMicroelectronic and foundry-supplied memory macros that are interpolated to 40$\,$\degree C, which is close to human body temperature.
We design each PE for its highest frequency, and scale the power when using them at lower frequency.
We run multi-corner, physically-aware synthesis, and use latency and power measurements from the worst variation corner. Table~\ref{tab:pe} shows these values. We taped out early designs of the PEs at 12$\,$nm to confirm these values.

\begin{table}[ht!]
\vspace{-2.5mm}
\caption{Latency and Power of the PEs.}
\centering
{\scriptsize
\begin{tabularx}{0.75\linewidth}{ lllll }
\toprule
\vspace{-1mm}Processing & Max Freq & \multicolumn{2}{l}{Power ($\mu W$)} & Latency\\\cmidrule(lr){3-4}
Elements & (MHz) & Leakage & Dyn/Elec &  (mS) \\
\midrule
FFT    & 15.7   & 141.97 & 9.02  & 4.00   \\
XCOR   & 85     & 377.00 & 44.11 & 4.00   \\
BBF    & 6      & 66.00  & 0.35  & 4.00   \\
SVM    & 3      & 99.00  & 0.53  & 1.67   \\
THR    & 16     & 2.00   & 0.11  & 0.06   \\
NEO    & 3      & 12.00  & 0.03  & 4.00   \\
HCONV  & 3      & 89.89  & 0.80  & 1.50   \\
NGRAM  & 0.2    & 15.69  & 0.08  & 1.50   \\
EMDH   & 0.03   & 10.47  & 0.00  & 0.04   \\
GATE   & 5      & 67.00  & 0.63  & 0.00   \\
HFREQ  & 2.88   & 61.98  & 0.52  & 4.00   \\
HCOMP  & 2.88   & 77.00  & 0.65  & 4.00   \\
NPACK  & 3      & 3.53   & 5.49  & 0.008  \\
UNPACK & 3      & 3.53   & 5.49  & 0.008  \\
DCOMP  & 16.393 & 7.20   & 0.14  & 0.50   \\
CCHECK & 16.393 & 7.20   & 0.14  & 0.50   \\
CSEL   & 0.1    & 4.00   & 6.00  & 0.04   \\
SC     & 3.2    & 95.30  & 1.64  & 0.03-4 \\
DTW    & 50     & 167.93 & 26.94 & 0.003  \\ 
\bottomrule
\end{tabularx}
}
\label{tab:pe}
\vspace{-2mm}
\end{table}

We assume that each node uses a standard 96-electrode array ~\cite{blackrock:utah} to sense neural activity, and a configurable 16-bit ADC~\cite{shen:sar} generating 30$\,$K samples per second per electrode. The ADC dissipates 2.88$\,$mW per sample from all 96 electrodes. Each node has a DAC to support electrical stimulation of brain tissue~\cite{medtronic:manual}, a process that consumes $\approx$0.6$\,$mW of power. 
\vspace{1mm}
\noindent {\bf Radio parameters.} We use a radio that can transmit/receive up to 10$\,$m to external devices, at 46$\,$Mbps, 250$\,$MHz frequency, and which consumes 9.2$\,$mW. For intra-\arch communication, we consider a state-of-the-art radio designed for safe implantation in the brain~\cite{rahmani:wirelessly}. While the radio was originally designed for asymmetric transmission/reception, we modify it for symmetric communication. Our intra-\arch radio supports a transmission distance of 20$\,$cm (i.e., $>$ 90$^{th}$ percentile head breadth~\cite{anthrop}). To estimate the power and data rates, we use path-loss models~\cite{molisch:report}, with a path-loss parameter of 3.5 for transmission through the brain, skull, and skin, consistent with prior studies~\cite{sarestoniemi:preliminary, taparugssanagorn:review}. We calculate that our radio can transmit/receive 7$\,$Mbps at 4.12$\,$GHz and consumes 1.721$\,$mW of power.

\begin{figure*}[h]
\centering
\subfloat[Maximum aggregate throughput of \arch versus alternative BCI architectures.]{
\centering
    \includegraphics[width=0.33\linewidth]{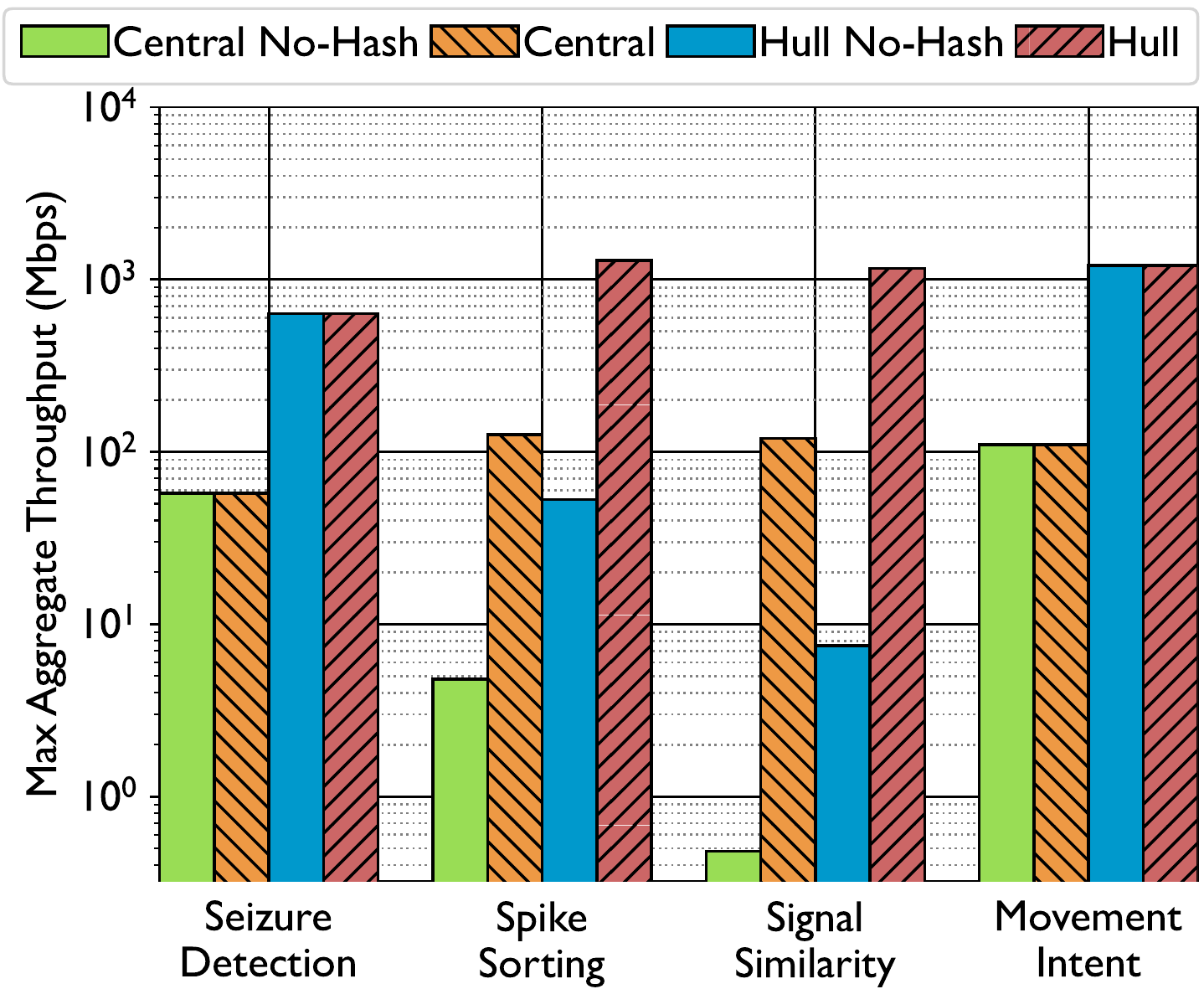}
    \label{fig:overallEval}
} \enspace
\subfloat[Maximum aggregate throughput of communication-dependent tasks in \arch.]{
\includegraphics[width=0.3\linewidth]{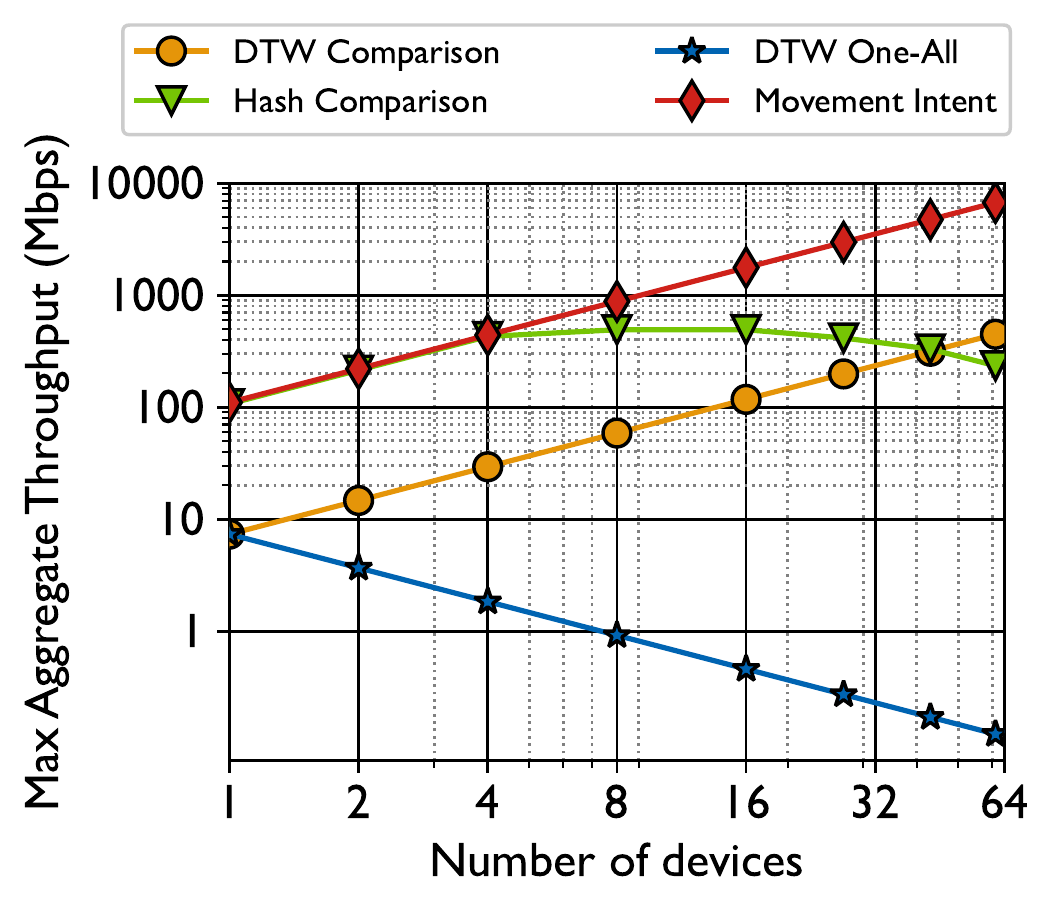}
\label{fig:comm}
} \enspace
\subfloat[Maximum throughput of tasks without inter-node communication, using re-designed PEs.]{
\includegraphics[width=0.3\linewidth]{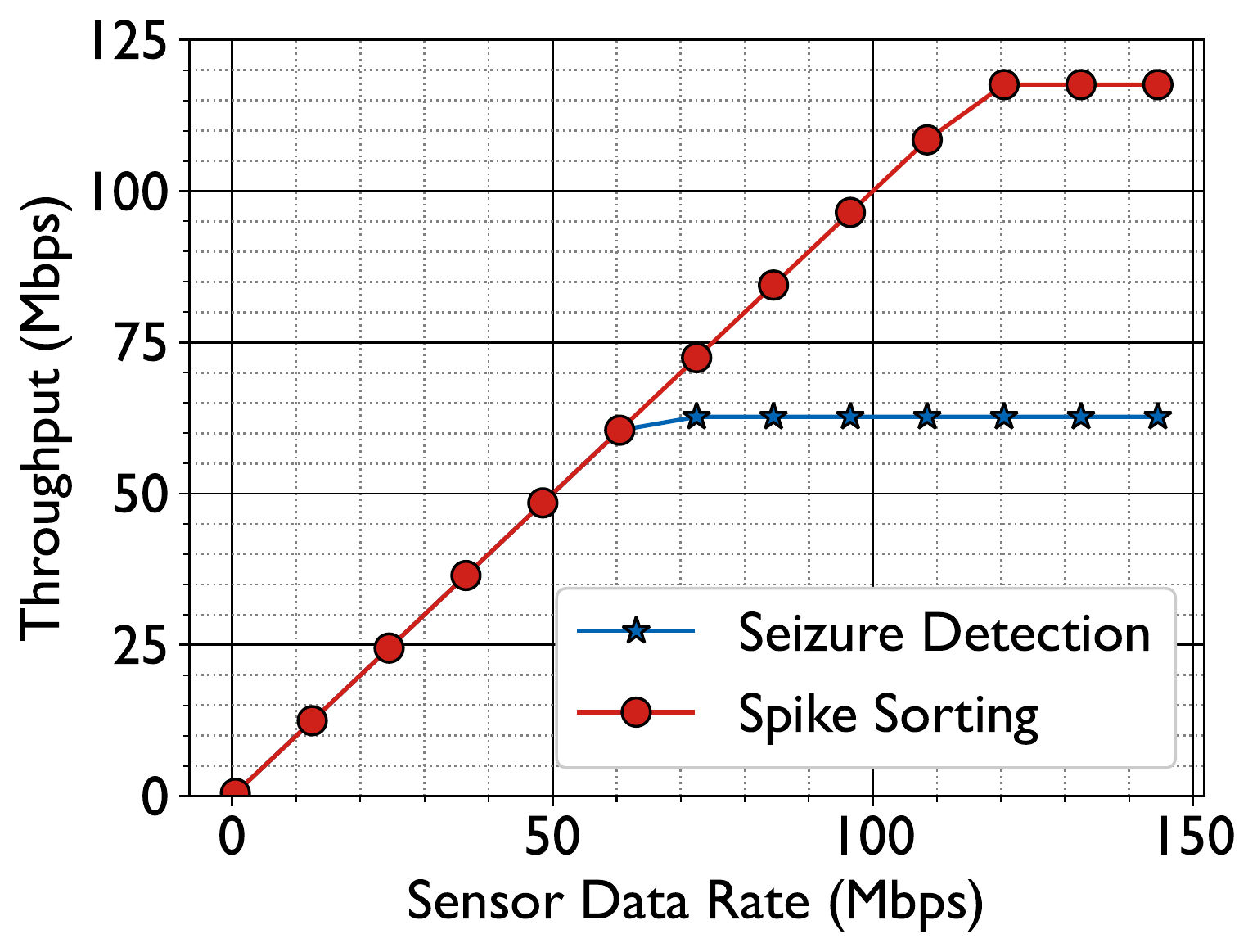}
\label{fig:ncommEval}
}
\caption{Experimental quantification of \arch's benefits.}
\vspace{-6mm}
\label{fig:eval}
\end{figure*} 


\vspace{1mm}
\noindent {\bf Non-volatile memory.} We use NVMs with 4$\,$KB page sizes and 1$\,$MB block sizes. The NVMs can read 8 bytes, write a page, or erase a block in one operation. We use SLC NAND parameters like erase time (1.5$\,$ms), program time (350$\,$us), and voltage (2.7$\,$V) from industrial technical reports~\cite{micron} with NVSim~\cite{dong_circuit_2012}. We choose a low operating power transistor type in NVSim, and use a temperature of 40$\,$\degree C.
NVSim assesses a leakage power of 0.252$\,$mW, dynamic energies of 164.4$\,$nJ and 261.143$\,$nJ per page for reads and writes, respectively. We also use these parameters to size our SC buffers to 24$\,$KB. 

\vspace{1mm}
\noindent {\bf Electrophysiological data.} We use publicly available electrophysiological data for our evaluation~\cite{spikeforest, ieeg}. For seizure detection and propagation, we use a data from the Mayo Clinic~\cite{ieeg} of a patient (label ``I001\_P013") with 76 electrodes implanted in the parietal/occipital lobes.
This data-set was recorded for 4 days at 5$\,$KHz, and is annotated with hundreds of seizure instances.
We upscaled the sampling frequency to 30$\,$KHz, and split the dataset to emulate multiple BCI devices.

We use consecutive and overlapping 4$\,$ms windows (120 samples) from the electrodes to detect seizures~\cite{shah_characterizing_2019}. For propagation, we check similarity with a seizure-positive signal in the last 100$\,$ms from electrode data in all nodes~\cite{shah_characterizing_2019}. For hash pipelines, we use one 8-bit hash for 120 sample data.

For spike sorting, we use the Spikeforest dataset~\cite{spikeforest, franklab}.
This dataset contains recordings collected from the CA1 region of a rat hippocampus using tetrode electrodes recorded at 30$\,$KHz sampling frequency. The dataset contains spikes from 10 neurons, with 65,$\,$000  spikes that were manually sorted.


\vspace{1mm}
\label{sub:compare}
\noindent {\bf Alternative system architectures.} Table~\ref{tab_compare} shows the systems that we compare \arch against. \textit{\arch No-Hash} uses the same \arch architecture but does not use hashes. The power saved by removing the hash processing PEs is allocated to the remaining tasks optimally. \textit{\arch No-Hash} does not require re-writing the applications for hash-based processing. \textit{Central} uses one processing node with the same processor as \arch, and multiple sensors that are connected using wires. Finally, \textit{Central No-Hash} is a centralized design without hash processing, like most existing BCIs~\cite{kassiri:closed, shupe_neurochip3_2021, ahmadi_towards_2019}. We do not consider wireless centralized designs as they need a radio and have lesser compute available than the wired ones. We also do not consider designs without memory as they do not support seizure propagation. We map our applications onto all systems using the ILP, ensuring that each node consumes $<$ 15$\,$mW.

\vspace{-2mm}
\begin{table}[h!]
\centering
{\scriptsize
\caption{Alternative BCI designs.}
\label{tab_compare}
\centering
\begin{tabulary}{0.75\columnwidth}{@{}p{0.2\columnwidth}p{0.16\columnwidth}p{0.18\columnwidth}p{0.18\columnwidth}@{}}
\toprule
Design & Architecture & Comparison & Communication \\
\toprule
\arch (Proposed) & Distributed & Hash, Signal & Wireless\\
\arch-No hash &   Distributed & Signal & Wireless\\
Central & Centralized & Hash, Signal & Wired \\
Central-No hash & Centralized & Signal & Wired\\
\bottomrule
\end{tabulary}
}
\vspace{-4mm}
\end{table}

\section{Evaluation}
\label{eval}

\subsection{Comparing BCI Architectures}
\label{sub:taskEval}

Figure~\ref{fig:overallEval} shows the maximum aggregate throughput of the systems in Table~\ref{tab_compare}. A task's maximum aggregate throughput is achieved when it is the only task running in the system, summed over all nodes. \textit{Central No-Hash} has the worst throughput for all tasks. This design suffers from having just one processor and from  using expensive signal processing. \textit{Central} increases throughput by an order of magnitude for tasks that benefit from hashing (spike sorting and signal similarity). However, the single processor remains the bottleneck for all tasks. 

\textit{\arch No-Hash} has distributed processors and enjoys higher aggregate seizure detection and movement intent. However, it performs poorly for tasks that need signal comparison (signal similarity, spike sorting). For these tasks, \textit{\arch No-Hash} has lower throughput than \textit{Central} because it does not use hashes. \arch uses distributed hash-based processing and has the highest aggregate throughput for all tasks. Compared to \textit{Central-No hash}, which is closest to state-of-the-art BCIs, \arch's data rates are an order of magnitude higher for seizure detection, and movement intent detection, and are nearly three orders of magnitude higher for signal similarity and spike sorting.

\subsection{Throughput for Communication-Dependent Tasks}
\label{sub:commEval}


Figure~\ref{fig:comm} shows the maximum aggregate throughput of the communication-dependent task (hash comparison, DTW comparison, and movement intent), with various node counts.
\textit{DTW Comparison} uses all-to-all comparison of raw signals. It has a lower throughput than the other tasks because only 16 out of 96 electrode signals can be transmitted for all-to-all comparison. The reason is that new electrode samples are obtained at 47$\,$Mbps from the ADC, but the intra-\arch radio can only transmit about 7$\,$Mbps.  Increasing the number of nodes decreases the throughput further because of the communication delays. Because \arch uses a TDMA network, where slots for network access are serialized, DTW Comparison has the worst throughput and scales poorly with node count. 

An alternative \textit{DTW One-All}, which only uses one-to-all DTW comparison, scales better since its communication latency doesn't increase with the number of nodes. However, a one-to-all comparison is insufficient for general BCI applications.

\textit{Hash Comparison} uses all-to-all hash communication to check for collisions. Its throughput increases to 470$\,$Mbps until 10 devices, after which it begins to decrease. When the number of nodes is small, few TDMA slots are required to exchange all hashes, enabling a linear increase in throughput as a function of node count. But, as node counts keep increasing, it takes longer to communicate all hashes and overall throughput reduces. 

Finally, \textit{Movement Intent} uses all-to-one communication of the partial SVM products. However, as the product is small, its throughput scales linearly with the number of nodes (note that the Y-axis in Figure~\ref{fig:comm} is logarithmic). It also has the highest aggregate throughput because it needs the least communication.

Figure~\ref{fig:comm} shows that hashing, and distributing the SVM computation in \arch enables it to scale to many regions and with higher data rates than what has been possible.


\subsection{Throughput for Non-Communicating Tasks}
\label{sub:ncommtaskEval}

We design our PEs for a maximum sensor rate of 47$\,$Mbps per node (Section~\ref{setup}). However, we study potential PE re-design to support higher processing rates for tasks that do not need communication. Figure~\ref{fig:ncommEval} shows the throughput of \textit{Seizure Detection} and \textit{Spike Sorting} for varying per-node signal sensor rates. Task throughput increases linearly up to 105$\,$Mbps for spike sorting, and 70$\,$Mbps for seizure detection. Beyond this sensing rate, the higher frequency of the PEs and ADCs results in exceeding the device power limit.  Nonetheless, these values are nearly twice as supported by existing single-implant BCIs and show the robustness of our methodology.


\subsection{Application Level Throughput}
\label{sub:seizEval}

\begin{figure}[h]
\vspace{-5mm}
\subfloat[Weighted throughput of seizure propagation tasks.]{
\centering
    \includegraphics[width=0.45\linewidth]{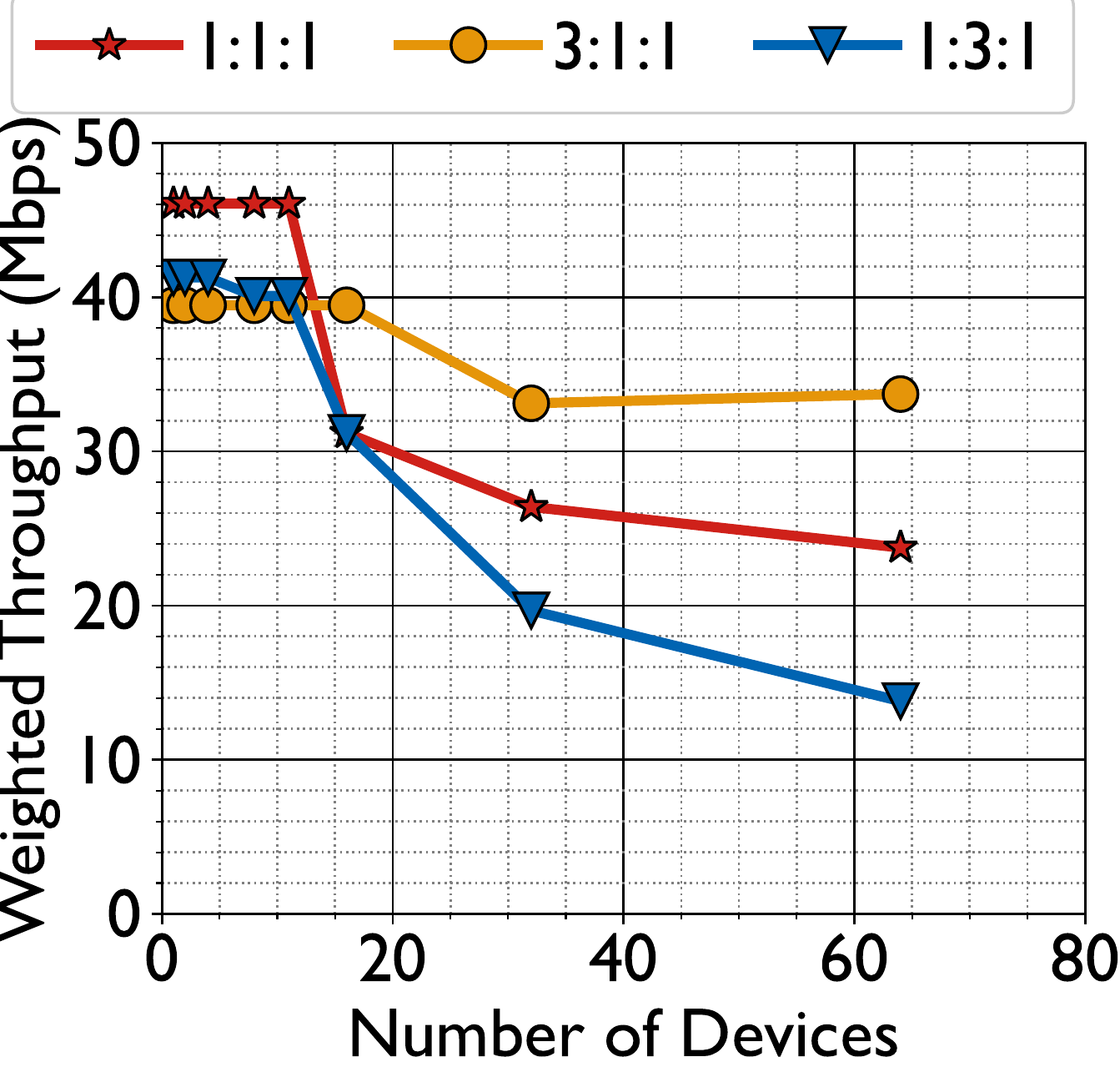}
    \label{fig:seizCost}
} \enspace
\subfloat[Movement intents per second (without device movement time).]{
\includegraphics[width=0.45\linewidth]{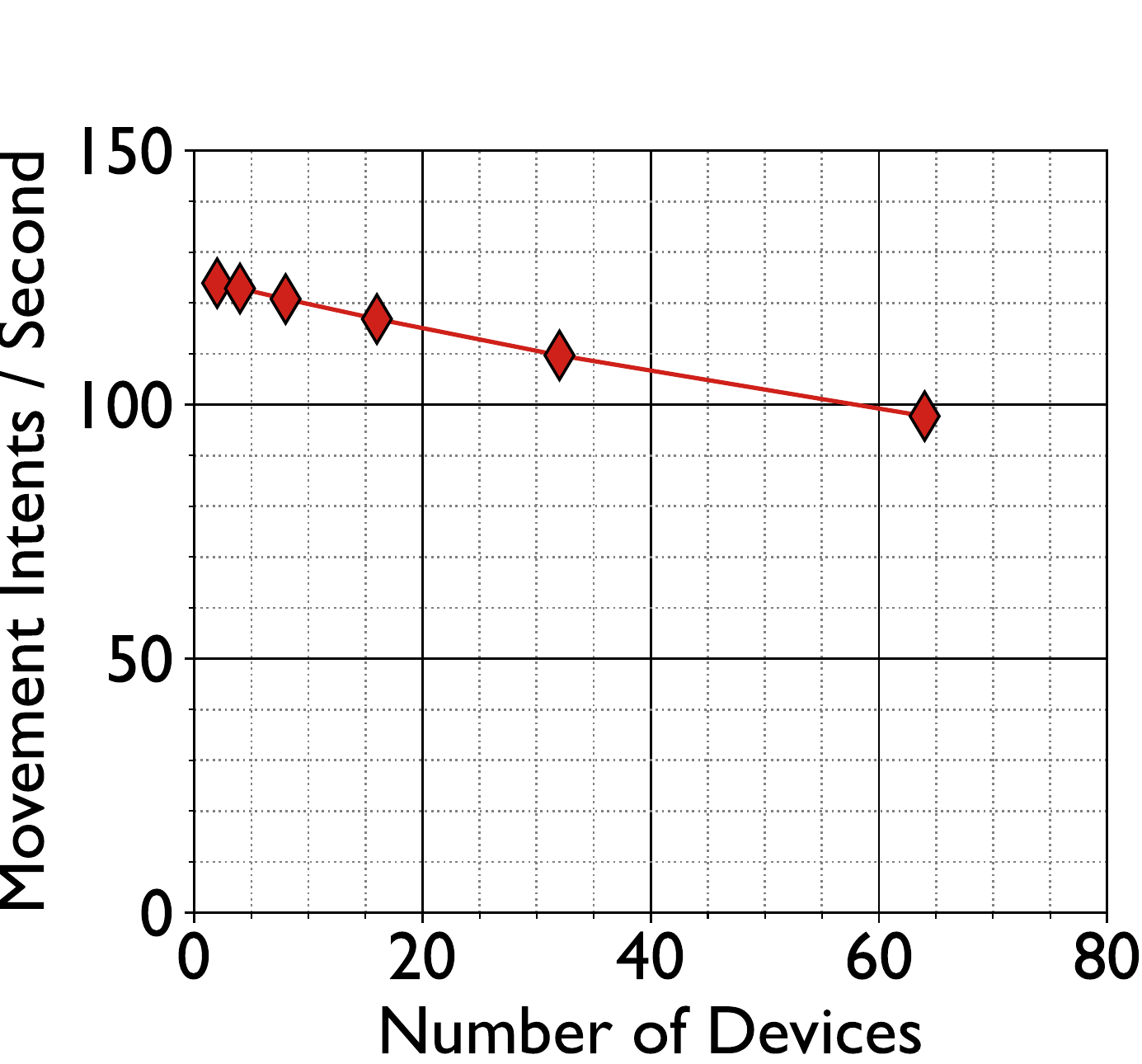}
\label{fig:intEvalomm}
}
\caption{Application level metrics on \arch.}
\vspace{-4mm}
\label{fig:app}
\end{figure}

The throughput achieved at the application level depends on the number of implanted nodes. Additionally, when there are multiple tasks, it depends on the priorities assigned to the application tasks. Recall that the ILP schedules applications to optimize a priority-weighted sum of the signals processed in each task. For seizure detection propagation, Figure~\ref{fig:seizCost} shows the weighted aggregate throughput as a function of the number of devices, for various weight choices (in the format seizure detection:hash comparison:DTW comparison). For an equal priority to seizure detection, DTW processing, and hash comparison, we find that the maximum throughput is achieved for 11 nodes. Other weight choices have different optimal node counts. Note that there is no comparable system for on-device seizure propagation---\arch is the first design with this feature.

Movement intent has only one task, and its throughput (in number of intents detected per second), is shown in Figure~\ref{fig:intEvalomm}. This metric accounts for only movement intent detection, and not for the variable response latency of the prosthetic device. 

\arch spike sorts up to 12,$\,$250 spikes per second per node with 82\% accuracy, comparing well to the state of the art~\cite{spikeforest}.

\subsection{Interactive Queries}
\label{sub:qeval}

We consider three types of common queries applied on data ranging from the past 100$\,$ms ($\approx$7$\,$MB over all nodes) to the past 1$\,$s ($\approx$60$\,$MB). They are: \textbf{Q1}, which returns all signals that were detected as a seizure; \textbf{Q2}, which returns all signals that matched with a template using a hash; and \textbf{Q3}, which returns all data in the timeframe.  For Q1 and Q2, we vary the fraction of data that tests positive for their condition.

\begin{figure}[h]
\vspace{-3mm}
    \centering
    \includegraphics[width=\linewidth]{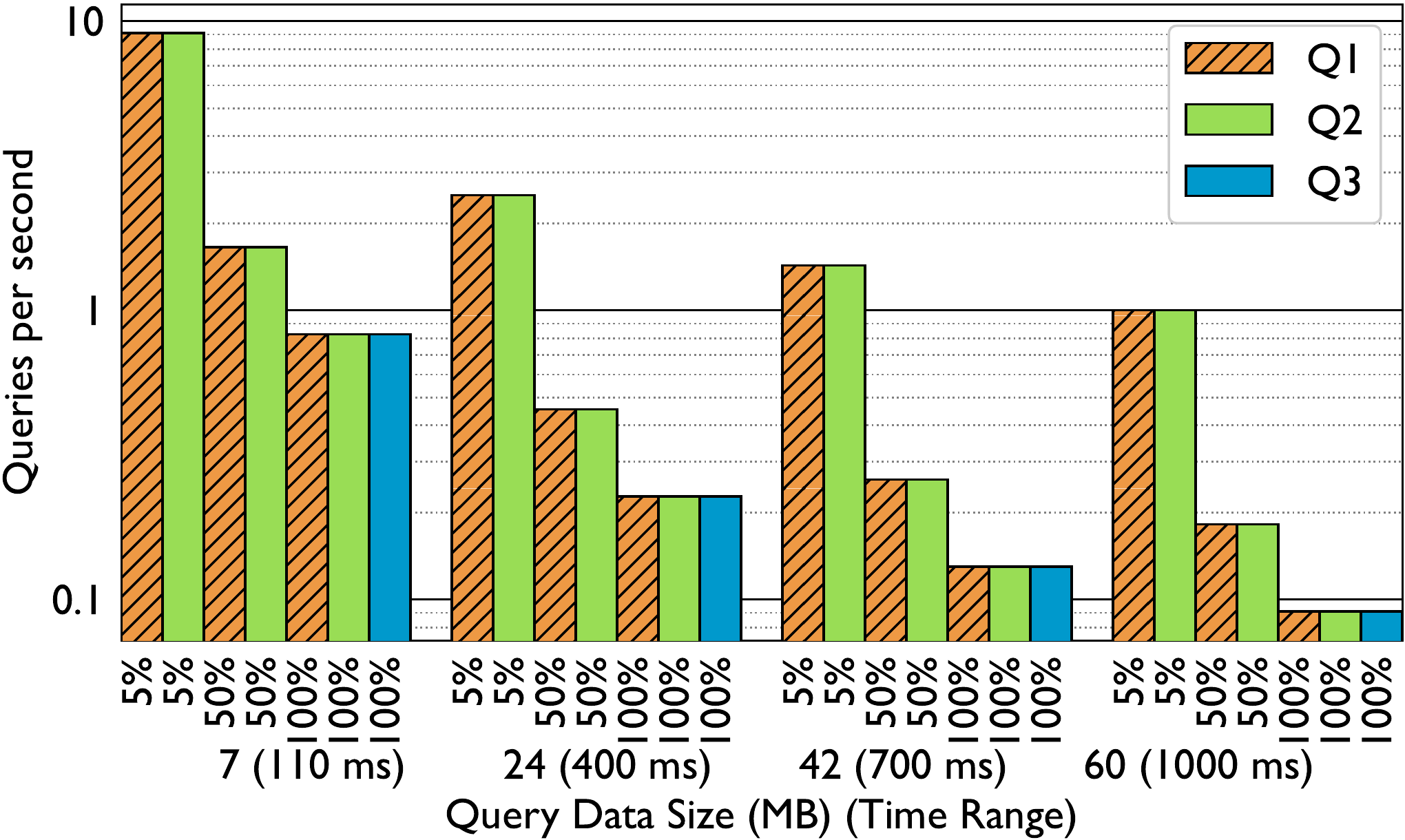}
    \caption{Interactive query throughput on \arch with 11 nodes.}
    \label{fig:qEval}
    \vspace{-3mm}
\end{figure}

Figure~\ref{fig:qEval} shows \arch's throughput with 11 nodes for our queries. \arch supports up to 10 queries per second (QPS) for Q1 and Q2 over the last 100$\,$ms data (the common case). If Q2 is run with DTW instead of hash-based search, we see a QPS of 8, which is only slightly lower, but the power consumption increases from 3.57$\,$mW for the hash vs the entire 15$\,$mW for DTW based matching. Thus, DTW-based matching is unsuitable when interactively querying in response to a  seizure. 

Q3 on this data takes 1.21$\,$s, yielding a throughput of $\approx$0.8. In interactive querying, the external radio, which consumes high power, is the bottleneck. 

As the data to be searched increases, the query latency increases linearly due to the radio latency. However, \arch can still process 1 QPS for Q1 and Q2 for the past 1$\,$s data ($\approx$60$\,$MB), making it suitable for real-time use.

\begin{figure*}[h]
\centering
           \begin{minipage}{.25\linewidth}
   \centering
     \includegraphics[width=\linewidth]{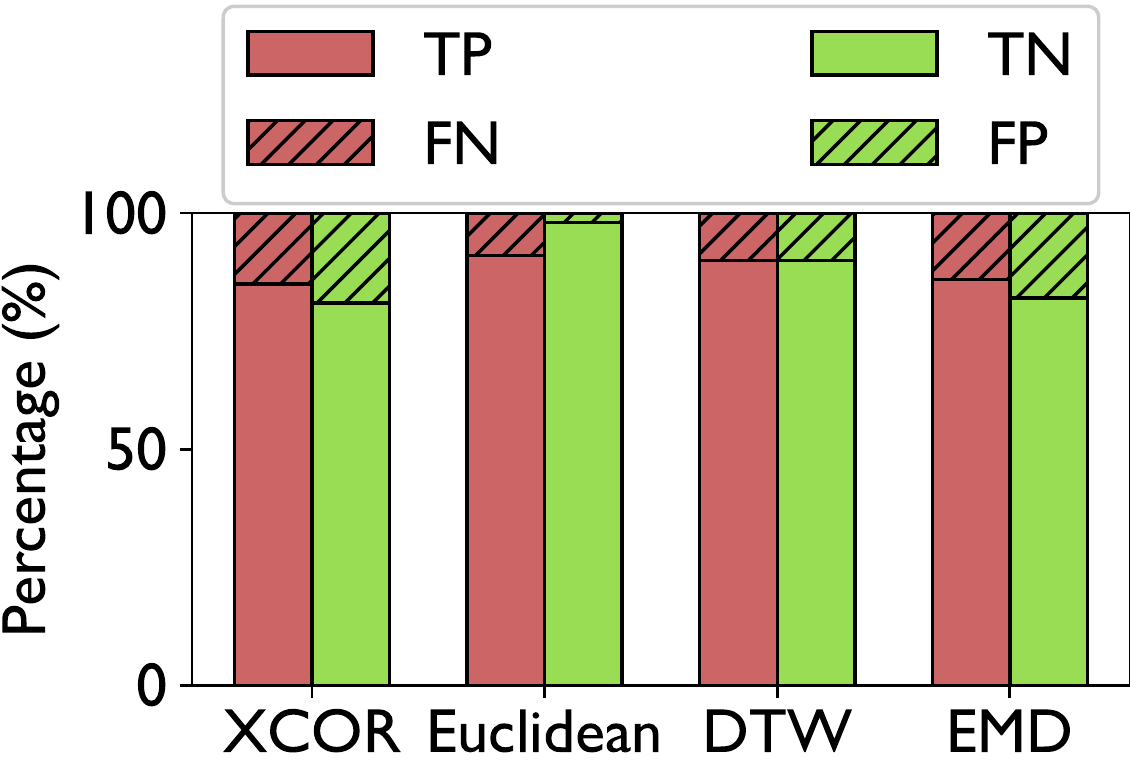}
    \caption{Hash accuracy.}
    \label{fig:hashEval2}
    \end{minipage}
    \begin{minipage}{.23\linewidth}
   \centering
     \includegraphics[width=\linewidth]{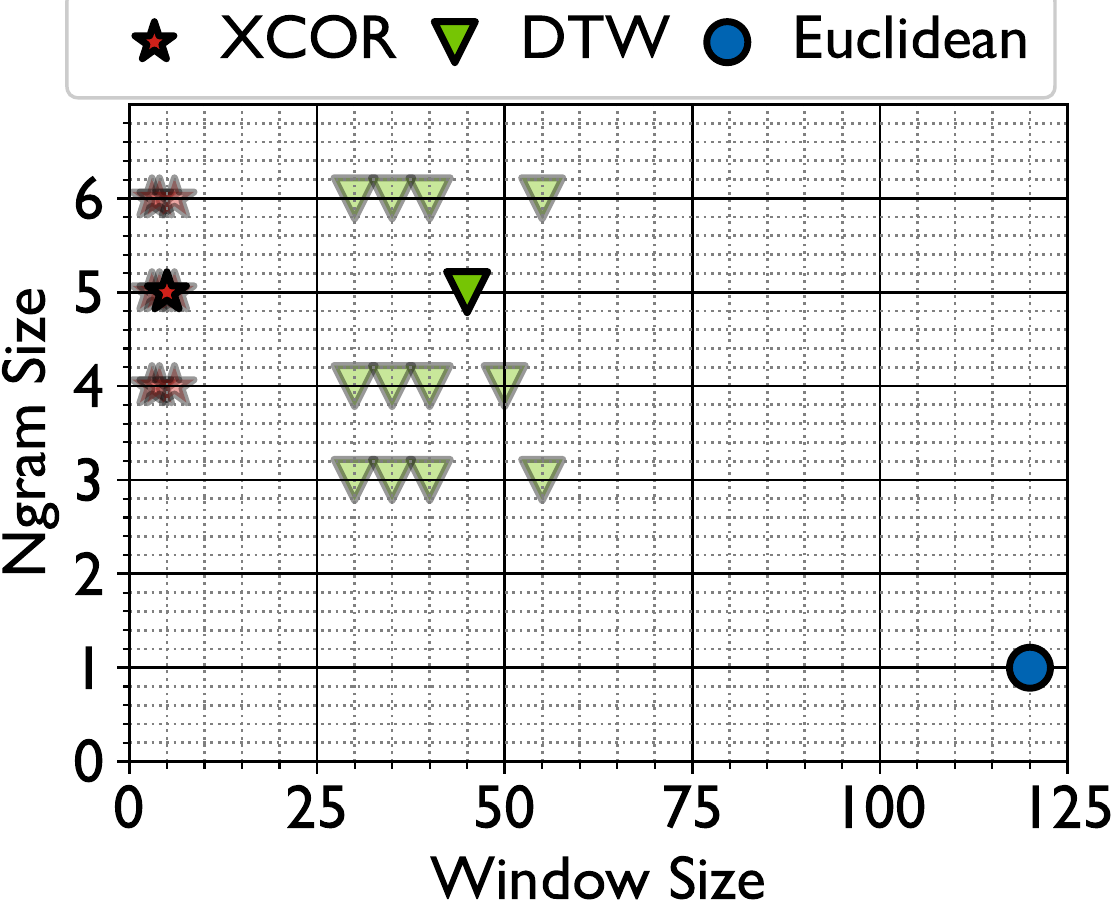}
    \caption{Hash flexibility.}
    \label{fig:hashEval}
    \end{minipage} \enspace
  \begin{minipage}{.2\linewidth}
   \centering
    \includegraphics[width=\linewidth]{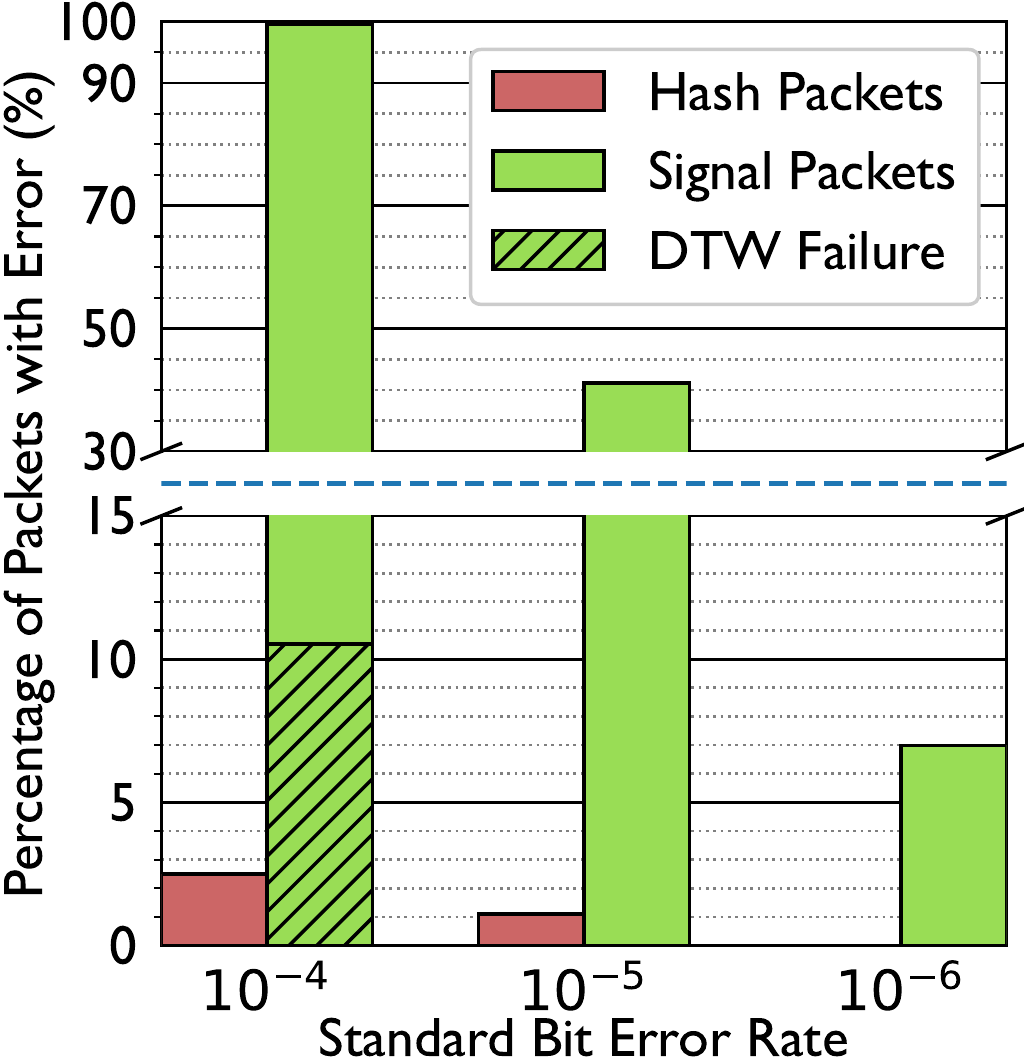}
    \caption{Bit error rates.}
\label{fig:bEval}
    \end{minipage}%
     \begin{minipage}{.27\linewidth}
   \centering
    \includegraphics[width=\linewidth]{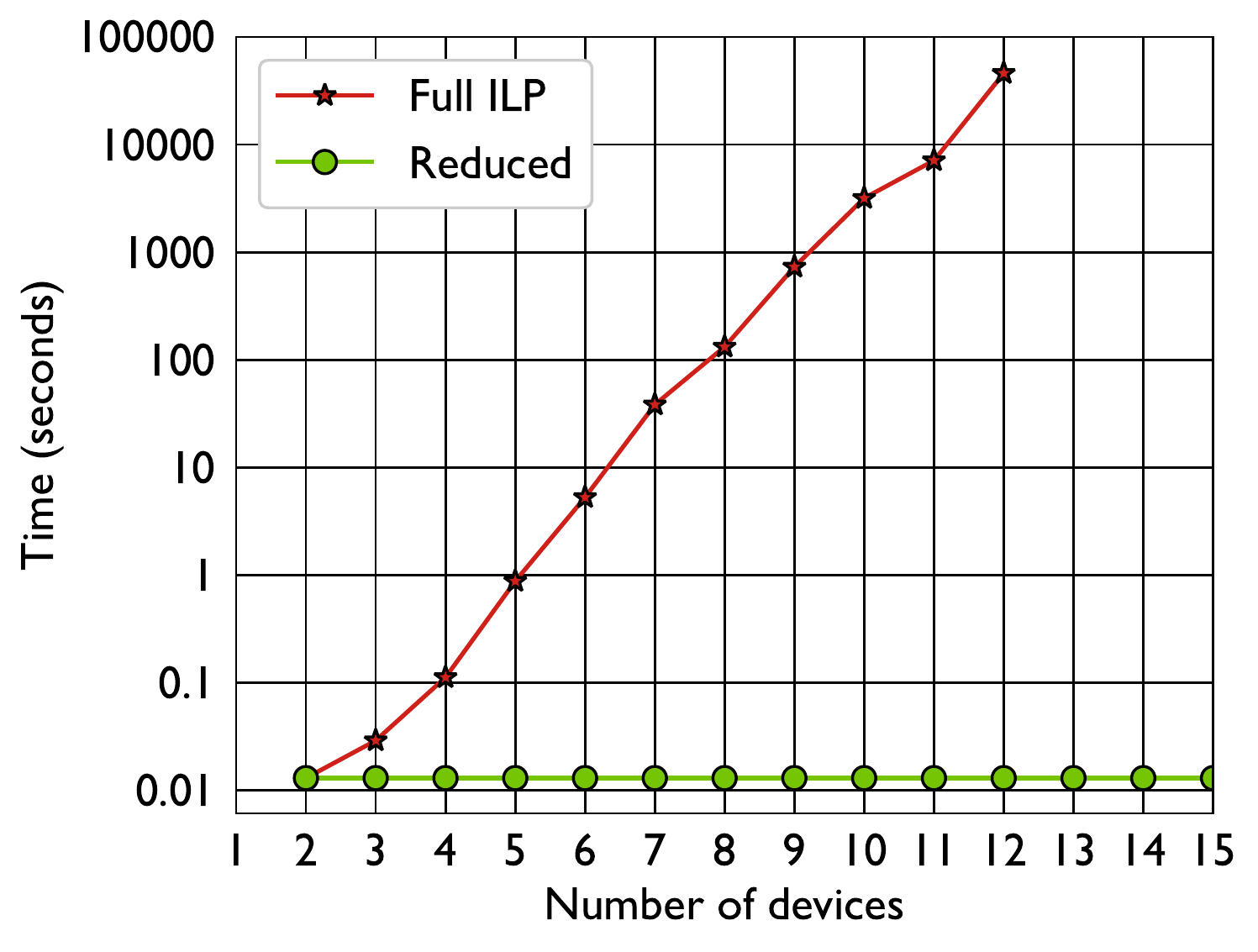}
    \caption{Time to solve the ILP.}
\label{fig:ilp}
    \end{minipage}%
\vspace{-3mm}
\end{figure*}

\subsection{Hashing}


\noindent
\textbf{Accuracy: } We vary the parameters of all our hash functions and show the performance of the best configuration for seizure propagation and spike sorting. Figure~\ref{fig:hashEval} shows the accuracy (TP: True positive, TN: True negative, FP: False positive, FN: False negative) for the four hash functions. XCOR and EMD hashes have $\approx$ 85\% accuracy while Euclidean and DTW have over 90\% accuracy. The high true positive rate of our DTW hash is particularly beneficial for the seizure propagation (note that false positives are removed using exact DTW).

\noindent
\textbf{Parameter selection: } Figure~\ref{fig:hashEval} shows the best parameters of our hash implementation (window size and n-gram size---Section~\ref{sub_hash}) to approximate each of Euclidean, cross correlation, and DTW similarity. We also show parameters (with lighter colors in the figure) that are within 90\% of the true positive rate achieved by the corresponding best configuration. This flexibility enables reusing a single fast hardware accelerator for different measures.

\subsection{Impact of Network Bit Error Rate}
\label{sec:eval:ber}

The intra-\arch network protocol drops packets carrying hashes when there is a checksum error but allows signal packets to flow into PEs since signal similarity measures are naturally resilient to a few errors.  We simulate various bit-error ratios (BERs) using uniformly random bit flips in the packet header and data.  
Figure~\ref{fig:bEval} shows the fraction of hash or signal packets with an error at different BERs, and the fraction of erroneous signal packets that flipped the similarity measure (DTW). For reference, the BER is \textless  $10^{-4}$  for the radio we use~\cite{rahmani:wirelessly}. 

Figure~\ref{fig:bEval} shows that signals and hashes suffer errors as BER increases, but signals are more susceptible since they are longer. But, 
even though several signal packets suffer errors, they have no impact on the final signal similarity outcome.

\subsection{ILP Performance}

The complexity of the ILP increases with the number of pipeline stages in the application and the number of \arch nodes. When all nodes are the same and have the same power/energy constraints, the schedule of one node can be replicated (with a constant offset) on all other nodes and remain optimal. We call this method \textit{Reduced} ILP. However, we cannot apply this method when the nodes are different or have different constraints. 
Figure \ref{fig:ilp} shows the time taken to solve the ILP and the reduced version for varying numbers of devices for the seizure propagation application. We measure this time when using GLPK, an open-source ILP solver, with default settings on an Intel-Xeon E5-2620 v3 machine with  93$\,$GB RAM. 
As expected, the solver time for the standard ILP increases exponentially with the number of devices, taking $\approx$2 hours with 11 devices. For $>$12 devices, the ILP did not finish within 24 hours and was terminated. The reduced ILP however, can be solved in less than 10ms for any number of devices.

\section{Related Work}


Commercial and research BCIs have focused largely on single brain location monitoring and stimulation~\cite{kassiri:closed, leary:nurip, medtronic:activapc, neuropace:rns, karageorgos:halo}, and have no support for distributed systems, making them inhospitable for the applications that we target.

Most implantable BCIs offer little to no storage capacity and stream data out continuously instead. NeuroChip~\cite{shupe_neurochip3_2021} is an exception, but is wired to an external case storing a 128 GB SD card that must be physically extracted for offline data analysis. \arch is the first to use storage for pre-processing and reduce computation by using the hash.

A growing interest in distributed analyses of the brain~\cite{jirsa_virtual_2017,bartolomei_defining_2017,andersen_exploring_2022, thakor_distributed_2021} has motivated the design of rudimentary multi-site BCIs~\cite{chung_high-density_2019, lee_neural_2021, ahmadi_towards_2019}.
Prior studies~\cite{lee_neural_2021, ahmadi_towards_2019} propose microchips that stream sensor data wirelessly to a central hub outside the skull using back-scattering radio techniques. Unfortunately, these approaches are restricted in their interfacing bandwidth as they rely on centralized processing and communication. 


Although recent work has studied unary neural networks on single-site BCIs \cite{wu_ubrain_2022}, we will study distributed neural network models for seizure detection, propagation, spike sorting, and movement intent for multi-side BCIs going forward. \arch can support any algorithm with linear computational complexity without significant changes to the ILP formulation.
However, neural network inference, which is super-linear, may require non-linear formulations for scheduling. Using MILP and approximations for such PEs may be a suitable extension.

\vspace{-1mm}
\section{Conclusion \& Future Work}

\arch  enables distributed BCI interfacing that can scale to multiple regions, and provides for the first time, on-device computation for important BCI applications. \arch offers two orders of magnitude higher task throughput, and real-time support for interactive querying with up to 10 QPS over 7$\,$ MB data or 1 QPS over 60$\,$MB data.

\arch will influence the wider field of IoT devices, ranging from low-power temperature and voltage sensors~\cite{upmu}, AR/VR devices, to devices in smart home, factory, and vehicle settings. These devices must collect and process large volumes of data on the edge, as communicating this data to centralized locations is likely to be near impossible for today's cloud infrastructure. Similar to \arch, a network of power-constrained devices will need to process large volumes of data, often with flexible processing requirements to support rapidly evolving use cases. \arch's design principles-- i.e., its modular PE architecture, fast-but-approximate hash-based approach to signal similarity, support for low-power and efficiently-indexed non-volatile storage, and a centralized planner that produces near-optimal mapping of task schedules to devices -- can be instrumental to success in other IoT environments as well.

\bibliographystyle{IEEEtran}
\bibliography{references}

\begin{thebibliography}{100}
\providecommand{\url}[1]{#1}
\csname url@samestyle\endcsname
\providecommand{\newblock}{\relax}
\providecommand{\bibinfo}[2]{#2}
\providecommand{\BIBentrySTDinterwordspacing}{\spaceskip=0pt\relax}
\providecommand{\BIBentryALTinterwordstretchfactor}{4}
\providecommand{\BIBentryALTinterwordspacing}{\spaceskip=\fontdimen2\font plus
\BIBentryALTinterwordstretchfactor\fontdimen3\font minus
  \fontdimen4\font\relax}
\providecommand{\BIBforeignlanguage}[2]{{%
\expandafter\ifx\csname l@#1\endcsname\relax
\typeout{** WARNING: IEEEtran.bst: No hyphenation pattern has been}%
\typeout{** loaded for the language `#1'. Using the pattern for}%
\typeout{** the default language instead.}%
\else
\language=\csname l@#1\endcsname
\fi
#2}}
\providecommand{\BIBdecl}{\relax}
\BIBdecl

\bibitem{andersen_exploring_2022}
R.~A. Andersen, T.~Aflalo, L.~Bashford, D.~Bj{\aa}nes, and S.~Kellis,
  ``{Exploring Cognition with Brain--Machine Interfaces},'' \emph{Annual Review
  of Psychology}, vol.~73, pp. 131--158, 2022.

\bibitem{lebedev_brain-machine_2017}
M.~A. Lebedev and M.~A. Nicolelis, ``{Brain-machine interfaces: From basic
  science to neuroprostheses and neurorehabilitation},'' \emph{Physiological
  reviews}, vol.~97, no.~2, pp. 767--837, 2017.

\bibitem{chandrasekaran_historical_2021}
S.~Chandrasekaran, M.~Fifer, S.~Bickel, L.~Osborn, J.~Herrero, B.~Christie,
  J.~Xu, R.~K. Murphy, S.~Singh, M.~F. Glasser \emph{et~al.}, ``Historical
  perspectives, challenges, and future directions of implantable brain-computer
  interfaces for sensorimotor applications,'' \emph{Bioelectronic medicine},
  vol.~7, no.~1, pp. 1--11, 2021.

\bibitem{widge_affective_2014}
A.~S. Widge, D.~D. Dougherty, and C.~T. Moritz, ``Affective brain-computer
  interfaces as enabling technology for responsive psychiatric stimulation,''
  \emph{Brain-Computer Interfaces}, vol.~1, no.~2, pp. 126--136, 2014.

\bibitem{shih_brain-computer_2012}
J.~J. Shih, D.~J. Krusienski, and J.~R. Wolpaw, ``Brain-computer interfaces in
  medicine,'' in \emph{Mayo clinic proceedings}, vol.~87, no.~3.\hskip 1em plus
  0.5em minus 0.4em\relax Elsevier, 2012, pp. 268--279.

\bibitem{mcfarland_therapeutic_2017}
D.~J. McFarland, J.~Daly, C.~Boulay, and M.~A. Parvaz, ``{Therapeutic
  applications of BCI technologies},'' \emph{Brain-Computer Interfaces},
  vol.~4, no. 1-2, pp. 37--52, 2017.

\bibitem{fazel-rezai_brain_2013}
L.~Huang and G.~van Luijtelaar, ``Brain computer interface for epilepsy
  treatment,'' \emph{Brain-Computer Interface Systems-Recent Progress and
  Future Prospects}, 2013.

\bibitem{muhl:survey}
C.~M{\"u}hl, B.~Allison, A.~Nijholt, and G.~Chanel, ``A survey of affective
  brain computer interfaces: principles, state-of-the-art, and challenges,''
  \emph{Brain-Computer Interfaces}, vol.~1, no.~2, pp. 66--84, 2014.

\bibitem{buzsaki:origin}
G.~Buzsáki, C.~A. Anastassiou, and C.~Koch, ``{The Origin of Extracellular
  Fields and Currents — EEG, ECoG, LFP and Spikes},'' \emph{Nature Reviews
  Neuroscience}, vol.~13, pp. 407--420, May 2012.

\bibitem{pesaran:investigating}
B.~Pesaran, M.~Vinck, G.~T. Einevoll, A.~Sirota, P.~Fries, M.~Siegel,
  W.~Truccolo, C.~E. Schroeder, and R.~Srinivasan, ``Investigating large-scale
  brain dynamics using field potential recordings: analysis and
  interpretation,'' \emph{Nature neuroscience}, vol.~21, no.~7, pp. 903--919,
  2018.

\bibitem{milan:invasive}
J.~del R.~Milan and J.~M. Carmena, ``{Invasive or Noninvasive: Understanding
  Brain-Machine Interface Technology [Conversations in BME]},'' \emph{IEEE
  Engineering in Medicine and Biology Magazine}, vol.~29, pp. 16--22, Jan 2010.

\bibitem{musk:integrated}
E.~Musk \emph{et~al.}, ``An integrated brain-machine interface platform with
  thousands of channels,'' \emph{Journal of medical Internet research},
  vol.~21, no.~10, p. e16194, 2019.

\bibitem{thakor_distributed_2021}
K.~M. Szostak, P.~Feng, F.~Mazza, and T.~G. Constandinou, ``{Distributed Neural
  Interfaces: Challenges and Trends in Scaling Implantable Technology},''
  \emph{Handbook of Neuroengineering}, pp. 1--37, 2021.

\bibitem{rapeaux_implantable_2021}
A.~B. Rapeaux and T.~G. Constandinou, ``Implantable brain machine interfaces:
  first-in-human studies, technology challenges and trends,'' \emph{Current
  opinion in biotechnology}, vol.~72, pp. 102--111, 2021.

\bibitem{fda_affairs_2021}
{U.S. Food and Drug Administration}, ``{FDA authorizes marketing of device to
  facilitate muscle rehabilitation in stroke patients},''
  \url{https://www.fda.gov/news-events/press-announcements/fda-authorizes-marketing-device-facilitate-muscle-rehabilitation-stroke-patients},
  April 2021.

\bibitem{neuropace:rns}
F.~T. Sun and M.~J. Morrell, ``{The RNS System: responsive cortical stimulation
  for the treatment of refractory partial epilepsy},'' \emph{Expert review of
  medical devices}, vol.~11, no.~6, pp. 563--572, 2014.

\bibitem{medtronic:activapc}
{Medtronic}, ``{Deep Brain Stimulation Systems - Activa PC},''
  \url{https://www.medtronic.com/us-en/healthcare-professionals/products/neurological/deep-brain-stimulation-systems/activa-pc.html},
  November 2018, {Retrieved August 10, 2019}.

\bibitem{fcc:new-guidance}
{U.S. Food and Drug Administration}, ``{Implanted Brain-Computer Interface
  (BCI) Devices for Patients with Paralysis or Amputation - Non-clinical
  Testing and Clinical Considerations},''
  \url{https://www.fda.gov/regulatory-information/search-fda-guidance-documents/implanted-brain-computer-interface-bci-devices-patients-paralysis-or-amputation-non-clinical-testing},
  February 2019, {Retrieved August 10, 2019}.

\bibitem{jason_synchron_2021}
J.~J. Han, ``{Synchron receives FDA approval to begin early feasibility study
  of their endovascular, brain-computer interface device},'' \emph{Artificial
  Organs}, vol.~45, no.~10, pp. 1134--1135, 2021.

\bibitem{serrano-amenos_thermal_2020}
C.~Serrano-Amenos, F.~Hu, P.~T. Wang, S.~Kellis, R.~A. Andersen, C.~Y. Liu,
  P.~Heydari, A.~H. Do, and Z.~Nenadic, ``Thermal analysis of a skull implant
  in brain-computer interfaces,'' in \emph{2020 42nd Annual International
  Conference of the IEEE Engineering in Medicine \& Biology Society
  (EMBC)}.\hskip 1em plus 0.5em minus 0.4em\relax IEEE, 2020, pp. 3066--3069.

\bibitem{wolf:thermal}
P.~D. Wolf, ``{Thermal Considerations for the Design of an Implanted Cortical
  Brain-Machine Interface (BMI)},'' \emph{Indwelling Neural Implants:
  Strategies for Contending with the In Vivo Environment}, 2008.

\bibitem{stevenson:advances}
I.~Stevenson and K.~Kording, ``{How Advances in Neural Recording Affect Data
  Analysis},'' \emph{Nature neuroscience}, vol.~14, pp. 139--42, 02 2011.

\bibitem{karageorgos:halo}
I.~Karageorgos, K.~Sriram, J.~Vesel{\`y}, M.~Wu, M.~Powell, D.~Borton,
  R.~Manohar, and A.~Bhattacharjee, ``Hardware-software co-design for
  brain-computer interfaces,'' in \emph{2020 ACM/IEEE 47th Annual International
  Symposium on Computer Architecture (ISCA)}.\hskip 1em plus 0.5em minus
  0.4em\relax IEEE, 2020, pp. 391--404.

\bibitem{zelmann_closes_2020}
R.~Zelmann, A.~C. Paulk, I.~Basu, A.~Sarma, A.~Yousefi, B.~Crocker,
  E.~Eskandar, Z.~Williams, G.~R. Cosgrove, D.~S. Weisholtz \emph{et~al.},
  ``{CLoSES: A platform for closed-loop intracranial stimulation in humans},''
  \emph{NeuroImage}, vol. 223, p. 117314, 2020.

\bibitem{jirsa_virtual_2017}
V.~K. Jirsa, T.~Proix, D.~Perdikis, M.~M. Woodman, H.~Wang,
  J.~Gonzalez-Martinez, C.~Bernard, C.~B{\'e}nar, M.~Guye, P.~Chauvel
  \emph{et~al.}, ``{The Virtual Epileptic Patient: Individualized whole-brain
  models of epilepsy spread},'' \emph{Neuroimage}, vol. 145, pp. 377--388,
  2017.

\bibitem{bartolomei_defining_2017}
F.~Bartolomei, S.~Lagarde, F.~Wendling, A.~McGonigal, V.~Jirsa, M.~Guye, and
  C.~B{\'e}nar, ``{Defining epileptogenic networks: Contribution of SEEG and
  signal analysis},'' \emph{Epilepsia}, vol.~58, no.~7, pp. 1131--1147, 2017.

\bibitem{kassiri:closed}
{Hossein Kassiri, Sana Tonekaboni, M. Tariqus Salam, Nima Soltani, karim
  Abdelhalim, Jose Luis Perez Velasquez, Roman Genov}, ``{Closed-Loop
  Neurostimulators: A Survey and A Seizure-Predicting Design Example for
  Intractable Epilepsy Treatment},'' \emph{IEEE Transactions on Biomedical
  Circuits and Systems}, vol.~11, no.~5, pp. 1026--1040, 2017.

\bibitem{leary:nurip}
{G. O’Leary and D. M. Groppe and T. A. Valiante and N. Verma and R. Genov},
  ``{Nurip: Neural interface processor for brain-state classification and
  programmable-waveform neurostimulation},'' \emph{IEEE Journal of Solid State
  Circuits}, vol.~53, 2018.

\bibitem{aziz:256:delta}
J.~N.~Y. Aziz, K.~Abdelhalim, R.~Shulyzki, R.~Genov, B.~L. Bardakjian,
  M.~Derchansky, D.~Serletis, and P.~L. Carlen, ``{256-Channel Neural Recording
  and Delta Compression Microsystem With 3D Electrodes},'' \emph{IEEE Journal
  of Solid-State Circuits}, vol.~44, no.~3, pp. 995--1005, March 2009.

\bibitem{chen:hardware}
T.~{Chen}, C.~{Jeng}, S.~{Chang}, H.~{Chiueh}, S.~{Liang}, Y.~{Hsu}, and
  T.~{Chien}, ``{A Hardware Implementation of Real-Time Epileptic Seizure
  Detector on FPGA},'' \emph{2011 IEEE Biomedical Circuits and Systems
  Conference (BioCAS)}, pp. 25--28, Nov 2011.

\bibitem{ahmadi_towards_2019}
N.~Ahmadi, M.~L. Cavuto, P.~Feng, L.~B. Leene, M.~Maslik, F.~Mazza,
  O.~Savolainen, K.~M. Szostak, C.-S. Bouganis, J.~Ekanayake \emph{et~al.},
  ``{Towards a Distributed, Chronically-Implantable Neural Interface},'' in
  \emph{2019 9th International IEEE/EMBS Conference on Neural Engineering
  (NER)}.\hskip 1em plus 0.5em minus 0.4em\relax IEEE, 2019, pp. 719--724.

\bibitem{lee_neural_2021}
J.~Lee, V.~Leung, A.-H. Lee, J.~Huang, P.~Asbeck, P.~P. Mercier,
  S.~Shellhammer, L.~Larson, F.~Laiwalla, and A.~Nurmikko, ``Neural recording
  and stimulation using wireless networks of microimplants,'' \emph{Nature
  Electronics}, vol.~4, no.~8, pp. 604--614, 2021.

\bibitem{zhu_closed-loop_2021}
B.~Zhu, U.~Shin, and M.~Shoaran, ``{Closed-Loop Neural Prostheses With On-Chip
  Intelligence: A Review and a Low-Latency Machine Learning Model for Brain
  State Detection},'' \emph{IEEE Transactions on Biomedical Circuits and
  Systems}, 2021.

\bibitem{zrenner_closed-loop_2016}
C.~Zrenner, P.~Belardinelli, F.~M{\"u}ller-Dahlhaus, and U.~Ziemann,
  ``{Closed-Loop Neuroscience and Non-Invasive Brain Stimulation: A Tale of Two
  Loop},'' \emph{Frontiers in cellular neuroscience}, vol.~10, p.~92, 2016.

\bibitem{translation_challenge}
M.~D. Murphy, D.~J. Guggenmos, D.~T. Bundy, and R.~J. Nudo, ``{{C}urrent
  {C}hallenges {F}acing the {T}ranslation of {B}rain {C}omputer {I}nterfaces
  from {P}reclinical {T}rials to {U}se in {H}uman {P}atients},'' \emph{Front
  Cell Neurosci}, vol.~9, p. 497, 2015.

\bibitem{hebb_creating_2014}
A.~O. Hebb, J.~J. Zhang, M.~H. Mahoor, C.~Tsiokos, C.~Matlack, H.~J. Chizeck,
  and N.~Pouratian, ``Creating the feedback loop: Closed loop
  neurostimulation,'' \emph{Neurosurgery Clinics of North America}, vol.~25,
  no.~1, pp. 187--204, 2014.

\bibitem{sladky_distributed_2022}
V.~Sladky, P.~Nejedly, F.~Mivalt, B.~H. Brinkmann, I.~Kim, E.~K. St.~Louis,
  N.~M. Gregg, B.~N. Lundstrom, C.~M. Crowe, T.~P. Attia \emph{et~al.},
  ``{Distributed brain co-processor for tracking spikes, seizures and behaviour
  during electrical brain stimulation},'' \emph{Brain Communications}, vol.~4,
  no.~3, p. fcac115, 2022.

\bibitem{kural_accurate_2022}
M.~A. Kural, J.~Jing, F.~F{\"u}rbass, H.~Perko, E.~Qerama, B.~Johnsen,
  S.~Fuchs, M.~B. Westover, and S.~Beniczky, ``{Accurate identification of EEG
  recordings with interictal epileptiform discharges using a hybrid approach:
  Artificial intelligence supervised by human experts},'' \emph{Epilepsia},
  vol.~63, no.~5, pp. 1064--1073, 2022.

\bibitem{simeral_home_2021}
J.~D. Simeral, T.~Hosman, J.~Saab, S.~N. Flesher, M.~Vilela, B.~Franco, J.~N.
  Kelemen, D.~M. Brandman, J.~G. Ciancibello, P.~G. Rezaii \emph{et~al.},
  ``{Home Use of a Percutaneous Wireless Intracortical Brain-Computer Interface
  by Individuals With Tetraplegia},'' \emph{{IEEE Transactions on Biomedical
  Engineering}}, vol.~68, no.~7, pp. 2313--2325, 2021.

\bibitem{yin_wireless_2014}
M.~Yin, D.~A. Borton, J.~Komar, N.~Agha, Y.~Lu, H.~Li, J.~Laurens, Y.~Lang,
  Q.~Li, C.~Bull \emph{et~al.}, ``{Wireless Neurosensor for Full-Spectrum
  Electrophysiology Recordings during Free Behavior},'' \emph{Neuron}, vol.~84,
  no.~6, pp. 1170--1182, 2014.

\bibitem{kim:thermal}
S.~M. Kim, P.~Tathireddy, R.~Normann, and F.~Solzbacher, ``{Thermal Impact of
  an Active 3-D Microelectrode Array Implanted in the Brain},'' \emph{IEEE
  Transactions on Neural Systems and Rehabilitation Engineering}, vol.~15, pp.
  493--501, Dec 2007.

\bibitem{sporns_graph_2018}
O.~Sporns, ``Graph theory methods: applications in brain networks,''
  \emph{Dialogues in clinical neuroscience}, 2022.

\bibitem{bilge_deep_2018}
M.~T. Bilge, A.~K. Gosai, and A.~S. Widge, ``{Deep Brain Stimulation in
  Psychiatry},'' \emph{Psychiatric Clinics}, vol.~41, no.~3, pp. 373--383,
  2018.

\bibitem{deco_great_2014}
G.~Deco and M.~L. Kringelbach, ``{Great Expectations: Using Whole-Brain
  Computational Connectomics for Understanding Neuropsychiatric Disorders},''
  \emph{Neuron}, vol.~84, no.~5, pp. 892--905, 2014.

\bibitem{gallego_going_2022}
J.~A. Gallego, T.~R. Makin, and S.~D. McDougle, ``Going beyond primary motor
  cortex to improve brain–computer interfaces,'' \emph{Trends in
  Neurosciences}, vol.~45, no.~3, pp. 176--183, 2022.

\bibitem{camargo-vargas_brain-computer_2021}
D.~Camargo-Vargas, M.~Callejas-Cuervo, and S.~Mazzoleni, ``{Brain-Computer
  Interfaces Systems for Upper and Lower Limb Rehabilitation: A Systematic
  Review},'' \emph{Sensors}, vol.~21, no.~13, p. 4312, 2021.

\bibitem{bensmaia_restoring_2014}
S.~J. Bensmaia and L.~E. Miller, ``Restoring sensorimotor function through
  intracortical interfaces: progress and looming challenges,'' \emph{Nature
  Reviews Neuroscience}, vol.~15, no.~5, pp. 313--325, 2014.

\bibitem{todorova_sort_2014}
S.~Todorova, P.~Sadtler, A.~Batista, S.~Chase, and V.~Ventura, ``To sort or not
  to sort: the impact of spike-sorting on neural decoding performance,''
  \emph{Journal of neural engineering}, vol.~11, no.~5, p. 056005, 2014.

\bibitem{rey_past_2015}
H.~G. Rey, C.~Pedreira, and R.~Q. Quiroga, ``Past, present and future of spike
  sorting techniques,'' \emph{Brain research bulletin}, vol. 119, pp. 106--117,
  2015.

\bibitem{shiao:svm}
H.~{Shiao}, V.~{Cherkassky}, J.~{Lee}, B.~{Veber}, E.~E. {Patterson}, B.~H.
  {Brinkmann}, and G.~A. {Worrell}, ``{SVM-Based System for Prediction of
  Epileptic Seizures From iEEG Signal},'' \emph{IEEE Transactions on Biomedical
  Engineering}, vol.~64, no.~5, pp. 1011--1022, May 2017.

\bibitem{niazi_detection_2011}
I.~K. Niazi, N.~Jiang, O.~Tiberghien, J.~F. Nielsen, K.~Dremstrup, and
  D.~Farina, ``Detection of movement intention from single-trial
  movement-related cortical potentials,'' \emph{Journal of Neural Engineering},
  vol.~8, no.~6, p. 066009, 2011.

\bibitem{jensen_detection_2014}
M.~Jochumsen, I.~K. Niazi, H.~Rovsing, C.~Rovsing, G.~A. Nielsen, T.~K.
  Andersen, N.~Dong, M.~E. S{\o}rensen, N.~Mrachacz-Kersting, N.~Jiang
  \emph{et~al.}, ``Detection of movement intentions through a single channel of
  electroencephalography,'' in \emph{Replace, Repair, Restore,
  Relieve--Bridging Clinical and Engineering Solutions in
  Neurorehabilitation}.\hskip 1em plus 0.5em minus 0.4em\relax Springer, 2014,
  pp. 465--472.

\bibitem{ojakangas_decoding_2006}
C.~L. Ojakangas, A.~Shaikhouni, G.~M. Friehs, A.~H. Caplan, M.~D. Serruya,
  M.~Saleh, D.~S. Morris, and J.~P. Donoghue, ``{Decoding Movement Intent From
  Human Premotor Cortex Neurons for Neural Prosthetic Applications},''
  \emph{Journal of Clinical Neurophysiology: Official Publication of the
  American Electroencephalographic Society}, vol.~23, no.~6, p. 577, 2006.

\bibitem{sternad_intention_2009}
J.~F. Kalaska, ``{From Intention to Action: Motor Cortex and the Control of
  Reaching Movements},'' \emph{Progress in Motor Control}, pp. 139--178, 2009,
  series Title: Advances in Experimental Medicine and Biology.

\bibitem{shakeel_review_2015}
A.~Shakeel, M.~S. Navid, M.~N. Anwar, S.~Mazhar, M.~Jochumsen, and I.~K. Niazi,
  ``{A Review of Techniques for Detection of Movement Intention Using
  Movement-Related Cortical Potentials},'' \emph{{Computational and
  Mathematical Methods in Medicine}}, vol. 2015, 2015.

\bibitem{taylor:direct}
D.~M. Taylor, S.~I.~H. Tillery, and A.~B. Schwartz, ``{Direct Cortical Control
  of 3D Neuroprosthetic Devices},'' \emph{Science}, vol. 296, no. 5574, pp.
  1829--1832, 2002.

\bibitem{fetz_restoring_2015}
E.~E. Fetz, ``Restoring motor function with bidirectional neural interfaces,''
  \emph{Progress in Brain Research}, vol. 218, pp. 241--252, 2015.

\bibitem{capogrosso_brainspine_2016}
M.~Capogrosso, T.~Milekovic, D.~Borton, F.~Wagner, E.~M. Moraud, J.-B.
  Mignardot, N.~Buse, J.~Gandar, Q.~Barraud, D.~Xing \emph{et~al.}, ``{A
  Brain--Spine Interface Alleviating Gait Deficits after Spinal Cord Injury in
  Primates},'' \emph{Nature}, vol. 539, no. 7628, pp. 284--288, 2016.

\bibitem{muller_linear_2003}
K.-R. Muller, C.~Anderson, and G.~Birch, ``Linear and nonlinear methods for
  brain-computer interfaces,'' \emph{IEEE Transactions on Neural Systems and
  Rehabilitation Engineering}, vol.~11, no.~2, pp. 165--169, 2003.

\bibitem{darie_delivering_2017}
R.~Darie, M.~Powell, and D.~Borton, ``{Delivering the Sense of Touch to the
  Human Brain},'' \emph{Neuron}, vol.~93, no.~4, pp. 728--730, 2017.

\bibitem{tafazoli_learning_2020}
S.~Tafazoli, C.~J. MacDowell, Z.~Che, K.~C. Letai, C.~R. Steinhardt, and T.~J.
  Buschman, ``Learning to control the brain through adaptive closed-loop
  patterned stimulation,'' \emph{Journal of Neural Engineering}, vol.~17,
  no.~5, p. 056007, 2020.

\bibitem{rutishauser_online_2006}
U.~Rutishauser, E.~M. Schuman, and A.~N. Mamelak, ``Online detection and
  sorting of extracellularly recorded action potentials in human medial
  temporal lobe recordings, in vivo,'' \emph{{Journal of Neuroscience
  Methods}}, vol. 154, no. 1-2, pp. 204--224, 2006.

\bibitem{sotomayor-gomez_spikeship_2020}
B.~Sotomayor-G{\'o}mez, F.~P. Battaglia, and M.~Vinck, ``{SpikeShip: A method
  for fast, unsupervised discovery of high-dimensional neural spiking
  patterns},'' \emph{bioRxiv}, pp. 2020--06, 2021.

\bibitem{grossberger_unsupervised_2018}
L.~Grossberger, F.~P. Battaglia, and M.~Vinck, ``{{U}nsupervised clustering of
  temporal patterns in high-dimensional neuronal ensembles using a novel
  dissimilarity measure},'' \emph{PLoS Comput Biol}, vol.~14, no.~7, p.
  e1006283, 07 2018.

\bibitem{Kim2017PhysiologicalTS}
Y.~B. Kim, ``{Physiological Time Series Retrieval and Prediction with
  Locality-Sensitive Hashing},'' Ph.D. dissertation, Massachusetts Institute of
  Technology, 2017.

\bibitem{cao_real-time_2016}
Y.~Cao, N.~Rakhilin, P.~H. Gordon, X.~Shen, and E.~C. Kan, ``A real-time spike
  classification method based on dynamic time warping for extracellular enteric
  neural recording with large waveform variability,'' \emph{{Journal of
  Neuroscience Methods}}, vol. 261, pp. 97--109, 2016.

\bibitem{luo_ssh_2016}
C.~Luo and A.~Shrivastava, ``{{SSH} (Sketch, Shingle, \& Hash) for Indexing
  Massive-Scale Time Series},'' in \emph{NIPS 2016 Time Series Workshop}.\hskip
  1em plus 0.5em minus 0.4em\relax PMLR, 2017, pp. 38--58.

\bibitem{gorisse_locality-sensitive_2012}
D.~Gorisse, M.~Cord, and F.~Precioso, ``{Locality-Sensitive Hashing for Chi2
  Distance},'' \emph{{{IEEE} Transactions on Pattern Analysis and Machine
  Intelligence}}, vol.~34, no.~2, pp. 402--409, 2011.

\bibitem{bhattacharjee_halo_2022}
A.~Bhattacharjee and R.~Manohar, ``{HALO: A Flexible and Low Power Processing
  Fabric for Brain-Computer Interfaces},'' in \emph{2022 IEEE Hot Chips 34
  Symposium (HCS)}.\hskip 1em plus 0.5em minus 0.4em\relax IEEE Computer
  Society, 2022, pp. 1--37.

\bibitem{karageorgos_balancing_2021}
I.~Karageorgos, K.~Sriram, J.~Veselý, N.~Lindsay, X.~Wen, M.~Wu, M.~Powell,
  D.~Borton, R.~Manohar, and A.~Bhattacharjee, ``Balancing specialized versus
  flexible computation in brain–computer interfaces,'' \emph{IEEE Micro},
  vol.~41, no.~3, pp. 87--94, 2021.

\bibitem{sun_closed-loop_2014}
F.~T. Sun and M.~J. Morrell, ``{Closed-loop Neurostimulation: The Clinical
  Experience},'' \emph{Neurotherapeutics}, vol.~11, no.~3, pp. 553--563, 2014.

\bibitem{vidal_review_2016}
G.~W.~V. Vidal, M.~L. Rynes, Z.~Kelliher, and S.~J. Goodwin, ``{Review of
  Brain-Machine Interfaces Used in Neural Prosthetics with New Perspective on
  Somatosensory Feedback through Method of Signal Breakdown},''
  \emph{Scientifica}, vol. 2016, 2016.

\bibitem{herff_potential_2020}
C.~Herff, D.~J. Krusienski, and P.~Kubben, ``{The Potential of
  Stereotactic-{EEG} for Brain-Computer Interfaces: Current Progress and Future
  Directions},'' \emph{Frontiers in Neuroscience}, vol.~14, p. 123, 2020.

\bibitem{consistenthash}
D.~Karger, E.~Lehman, T.~Leighton, R.~Panigrahy, M.~Levine, and D.~Lewin,
  ``{Consistent Hashing and Random Trees: Distributed Caching Protocols for
  Relieving Hot Spots on the World Wide Web},'' in \emph{ACM Symposium on
  Theory of Computing (STOC)}, 1997.

\bibitem{dtw}
J.~Kruskall and M.~Liberman, ``The symmetric time warping algorithm: From
  continuous to discrete,'' \emph{{Time Warps, String Edits and Macromolecules:
  The Theory and Practice of Sequence Comparison}}, 1983.

\bibitem{sakoe:dynamic}
H.~Sakoe and S.~Chiba, ``Dynamic programming algorithm optimization for spoken
  word recognition,'' \emph{IEEE Transactions on Acoustics, Speech, and Signal
  Processing}, vol.~26, pp. 159--165, 1978.

\bibitem{pele:fastemd}
O.~Pele and M.~Werman, ``{Fast and robust Earth Mover's Distances},'' in
  \emph{2009 IEEE 12th international conference on computer vision}.\hskip 1em
  plus 0.5em minus 0.4em\relax IEEE, 2009, pp. 460--467.

\bibitem{pu_fundamental_2005}
I.~M. Pu, \emph{Fundamental data compression}.\hskip 1em plus 0.5em minus
  0.4em\relax Butterworth-Heinemann, 2005.

\bibitem{elias_code}
P.~Elias, ``Universal codeword sets and representations of the integers,''
  \emph{IEEE Transactions on Information Theory}, vol.~21, no.~2, pp. 194--203,
  1975.

\bibitem{bingham2018qdi}
N.~Bingham and R.~Manohar, ``Qdi constant-time counters,'' \emph{IEEE
  Transactions on Very Large Scale Integration (VLSI) Systems}, vol.~27, no.~1,
  pp. 83--91, 2018.

\bibitem{crc}
W.~W. Peterson and D.~T. Brown, ``Cyclic codes for error detection,''
  \emph{Proceedings of the IRE}, vol.~49, no.~1, pp. 228--235, 1961.

\bibitem{yun:pausible}
K.~Yun and R.~Donohue, ``Pausible clocking: a first step toward heterogeneous
  systems,'' in \emph{Proceedings International Conference on Computer Design.
  VLSI in Computers and Processors}, 1996, pp. 118--123.

\bibitem{moreno:synthesis}
A.~Moreno and J.~Cortadella, ``Synthesis of all-digital delay lines,'' in
  \emph{2017 23rd IEEE International Symposium on Asynchronous Circuits and
  Systems (ASYNC)}, 2017, pp. 75--82.

\bibitem{mills_1995}
D.~Mills, ``{Simple Network Time Protocol (SNTP)},'' RFC Editor, {RFC} 1769,
  1995.

\bibitem{sql}
D.~D. Chamberlin and R.~F. Boyce, ``{SEQUEL: A structured English query
  language},'' in \emph{{Proceedings of the 1974 ACM SIGFIDET (now SIGMOD)
  workshop on Data description, access and control}}.\hskip 1em plus 0.5em
  minus 0.4em\relax Association for Computing Machinery, 1974, pp. 249--264.

\bibitem{blackrock:utah}
{Blackrock Microsystems}, ``{The Benchmark for Multichannel, High-density
  Neural Recording},''
  \url{https://www.blackrockmicro.com/electrode-types/utah-array/}, {Retrieved
  August 10, 2019}.

\bibitem{shen:sar}
J.~Shen, A.~Shikata, L.~D. Fernando, N.~Guthrie, B.~Chen, M.~Maddox,
  N.~Mascarenhas, R.~Kapusta, and M.~C.~W. Coln, ``A 16-bit 16-ms/s sar adc
  with on-chip calibration in 55-nm cmos,'' \emph{IEEE Journal of Solid-State
  Circuits}, vol.~53, no.~4, pp. 1149--1160, 2018.

\bibitem{medtronic:manual}
{Medtronic}, ``{Medtronic Activa PC Multi-program neurostimulator implant
  manual},''
  \url{http://www.neuromodulation.ch/sites/default/files/pictures/activa_PC_DBS_implant_manuel.pdf},
  2008, {Retrieved August 10, 2019}.

\bibitem{rahmani:wirelessly}
H.~Rahmani and A.~Babakhani, ``A wirelessly powered reconfigurable fdd radio
  with on-chip antennas for multi-site neural interfaces,'' \emph{IEEE Journal
  of Solid-State Circuits}, vol.~56, no.~10, pp. 3177--3190, 2021.

\bibitem{anthrop}
J.~W. Young, ``{Head and Face Anthropometry of Adult U.S. Citizens},'' {Federal
  Aviation Administration}, Tech. Rep., 1993.

\bibitem{molisch:report}
A.~F. Molisch, K.~Balakrishnan, C.-C. Chong, S.~Emami, A.~Fort, J.~Karedal,
  J.~Kunisch, H.~Schantz, U.~Schuster, and K.~Siwiak, ``{IEEE 802.15. 4a
  channel model-final report},'' \emph{IEEE P802}, vol.~15, no.~04, p. 0662,
  2004.

\bibitem{sarestoniemi:preliminary}
M.~S{\"a}rest{\"o}niemi, C.~Pomalaza-Raez, K.~Sayrafian, T.~Myllyl{\"a}, and
  J.~Iinatti, ``{A Preliminary Study of RF Propagation for High Data Rate Brain
  Telemetry},'' in \emph{{Body Area Networks. Smart IoT and Big Data for
  Intelligent Health Management}}.\hskip 1em plus 0.5em minus 0.4em\relax Cham:
  Springer International Publishing, 2022, pp. 126--138.

\bibitem{taparugssanagorn:review}
A.~Taparugssanagorn, A.~Rabbachin, M.~H{\"a}m{\"a}l{\"a}inen, J.~Saloranta,
  J.~Iinatti \emph{et~al.}, ``{A Review of Channel Modelling for Wireless Body
  Area Network in Wireless Medical Communications},'' \emph{{The 11th
  International Symposium on Wireless Personal Multimedia Communications
  (WPMC}}, 2008.

\bibitem{micron}
``{MT29F128G08AKCABH2-10},''
  \url{https://www.micron.com/products/nand-flash/slc-nand/part-catalog/mt29f128g08akcabh2-10},
  {Retrieved November 17, 2022}.

\bibitem{dong_circuit_2012}
X.~Dong, C.~Xu, Y.~Xie, and N.~P. Jouppi, ``Nvsim: A circuit-level performance,
  energy, and area model for emerging nonvolatile memory,'' \emph{IEEE
  Transactions on Computer-Aided Design of Integrated Circuits and Systems},
  vol.~31, no.~7, pp. 994--1007, 2012.

\bibitem{spikeforest}
J.~Magland, J.~J. Jun, E.~Lovero, A.~J. Morley, C.~L. Hurwitz, A.~P. Buccino,
  S.~Garcia, and A.~H. Barnett, ``{SpikeForest, reproducible web-facing
  ground-truth validation of automated neural spike sorters},'' \emph{Elife},
  vol.~9, p. e55167, 2020.

\bibitem{ieeg}
``{IEEG.ORG},'' \url{https://www.ieeg.org}, {Retrieved November 17, 2022}.

\bibitem{shah_characterizing_2019}
P.~Shah, A.~Ashourvan, F.~Mikhail, A.~Pines, L.~Kini, K.~Oechsel, S.~R. Das,
  J.~M. Stein, R.~T. Shinohara, D.~S. Bassett \emph{et~al.}, ``Characterizing
  the role of the structural connectome in seizure dynamics,'' \emph{Brain},
  vol. 142, no.~7, pp. 1955--1972, 2019.

\bibitem{franklab}
J.~E. Chung, J.~F. Magland, A.~H. Barnett, V.~M. Tolosa, A.~C. Tooker, K.~Y.
  Lee, K.~G. Shah, S.~H. Felix, L.~M. Frank, and L.~F. Greengard, ``{A Fully
  Automated Approach to Spike Sorting},'' \emph{Neuron}, vol.~95, no.~6, pp.
  1381--1394, 2017.

\bibitem{shupe_neurochip3_2021}
L.~E. Shupe, F.~P. Miles, G.~Jones, R.~Yun, J.~Mishler, I.~Rembado, R.~L.
  Murphy, S.~I. Perlmutter, and E.~E. Fetz, ``{Neurochip3: An Autonomous
  Multichannel Bidirectional Brain-Computer Interface for Closed-Loop
  Activity-Dependent Stimulation},'' \emph{{Frontiers in Neuroscience}},
  vol.~15, 2021.

\bibitem{chung_high-density_2019}
J.~E. Chung, H.~R. Joo, J.~L. Fan, D.~F. Liu, A.~H. Barnett, S.~Chen,
  C.~Geaghan-Breiner, M.~P. Karlsson, M.~Karlsson, K.~Y. Lee \emph{et~al.},
  ``{High-Density, Long-Lasting, and Multi-region Electrophysiological
  Recordings Using Polymer Electrode Arrays},'' \emph{Neuron}, vol. 101, no.~1,
  pp. 21--31, 2019.

\bibitem{wu_ubrain_2022}
D.~Wu, J.~Li, Z.~Pan, Y.~Kim, and J.~S. Miguel, ``{uBrain: A Unary Brain
  Computer Interface},'' in \emph{Proceedings of the 49th Annual International
  Symposium on Computer Architecture}.\hskip 1em plus 0.5em minus 0.4em\relax
  Association for Computing Machinery, 2022, pp. 468--481.

\bibitem{upmu}
E.~M. Stewart, A.~Liao, and C.~Roberts, ``Open $\mu$pmu: A real world reference
  distribution micro-phasor measurement unit data set for research and
  application development,'' \emph{IEEE Power Engineering Letters}, 2016.

\end{thebibliography}

\end{document}